\pdfoutput=1
\documentclass[twocolumn]{aastex63}

\usepackage{amsmath}
\usepackage{amssymb}
\usepackage{multirow}
\usepackage{xspace}

\newcommand{\aumic}{\object{AU Mic}\xspace}

\newcommand{\tess}{\textit{TESS}\xspace}
\newcommand{\spitzer}{{\sl Spitzer}\xspace}

\newcommand{\exofast}{{\tt EXOFASTv2}\xspace}
\newcommand{\irac}{{\sl IRAC}\xspace}

\newcommand{\tbnm}{\tablenotemark}
\newcommand{\magv}{m$_{\mbox{\scriptsize V}}$}
\newcommand{\magt}{m$_{\mbox{\scriptsize TESS}}$}
\newcommand{\alphaj}{$\alpha_{\mbox{\scriptsize J2000}}$\xspace}
\newcommand{\deltaj}{$\delta_{\mbox{\scriptsize J2000}}$\xspace}
\newcommand{\pmalpha}{$\mu_{\mbox{\scriptsize $\alpha$}}$\xspace}
\newcommand{\pmdelta}{$\mu_{\mbox{\scriptsize $\delta$}}$\xspace}
\newcommand{\mstar}{M$_{\star}$\xspace}
\newcommand{\rstar}{R$_{\star}$\xspace}
\newcommand{\lstar}{L$_{\star}$\xspace}
\newcommand{\msun}{M$_{\sun}$\xspace}
\newcommand{\rsun}{R$_{\sun}$\xspace}
\newcommand{\lsun}{L$_{\sun}$\xspace}
\newcommand{\mplan}{M$_{\mbox{\scriptsize p}}$\xspace}
\newcommand{\rplan}{R$_{\mbox{\scriptsize p}}$\xspace}
\newcommand{\mjup}{M$_{\mbox{\scriptsize J}}$\xspace}
\newcommand{\rjup}{R$_{\mbox{\scriptsize J}}$\xspace}
\newcommand{\ljup}{L$_{\mbox{\scriptsize J}}$\xspace}
\newcommand{\mear}{M$_{\earth}$\xspace}
\newcommand{\rear}{R$_{\earth}$\xspace}
\newcommand{\teff}{T$_{\mbox{\scriptsize eff}}$\xspace}
\newcommand{\prot}{P$_{\mbox{\scriptsize rot}}$\xspace}
\newcommand{\porb}{P$_{\mbox{\scriptsize orb}}$\xspace}
\newcommand{\tc}{T$_{\mbox{\scriptsize C}}$\xspace}
\newcommand{\tdur}{$\tau_{14}$\xspace}
\newcommand{\logmstmsu}{log$_{10}\left(\frac{\mbox{\scriptsize M}_{\scriptsize \star}}{\mbox{\scriptsize M}_{\mbox{\scriptsize $\sun$}}}\right)$\xspace}

\shortauthors{Collins et al.}

\begin{document}

\title{Thermal Eclipse Observation of the Young Hot Neptune AU Mic b with Spitzer}

\correspondingauthor{Kevin I. Collins}
\email{kcolli3@gmu.edu}

\author[0000-0003-2781-3207]{Kevin I. Collins}
\affiliation{George Mason University, 4400 University Drive
Fairfax, VA 22030, USA}

\author[0000-0002-8864-1667]{Peter Plavchan}
\affiliation{George Mason University, 4400 University Drive
Fairfax, VA 22030, USA}

\author[0000-0002-3321-4924]{Zachory Berta-Thompson}
\affiliation{Department of Astrophysical and Planetary Sciences, University of Colorado, 2000 Colorado Avenue, Boulder, CO 80309, USA}

\author[0000-0002-1013-2811]{Christoph Mordasini}
\affiliation{Space Research and Planetary Sciences, Physics Institute, University of Bern, Gesellschaftsstrasse 6, 3012 Bern, Switzerland}

\author[0000-0001-8832-4488]{Dan Huber}
\affiliation{Institute for Astronomy, University of Hawai‘i, 2680 Woodlawn Drive, Honolulu, HI 96822, USA}

\author[0000-0002-4818-7885]{Jamie Tayar}
\affiliation{Department of Astronomy, University of Florida, Bryant Space Science Center, Stadium Road, Gainesville, FL 32611, USA}

\author[0000-0002-9355-5165]{Brice-Olivier Demory}
\affiliation{Center for Space and Habitability, University of Bern, Gesellschaftsstrasse 6, 3012 Bern, Switzerland}

\author[0000-0002-0583-0949]{Ward S. Howard}
\affiliation{Department of Astrophysical and Planetary Sciences, University of Colorado, 2000 Colorado Avenue, Boulder, CO 80309, USA}

\author{Nicholas Law}
\affiliation{Department of Physics and Astronomy, The University of North Carolina at Chapel Hill, Chapel Hill, NC 27599-3255, USA}

\author[0000-0001-7139-2724]{Thomas Barclay}
\affiliation{NASA Goddard Space Flight Center, 8800 Greenbelt Road, Greenbelt, MD 20771, USA}

\author[0000-0002-1835-1891]{Ian J. M. Crossfield}
\affiliation{Department of Physics \& Astronomy, University of Kansas, 1251 Wescoe Hall Dr, Lawrence, KS 66045, USA}

\author[0000-0003-2313-467X]{Diana Dragomir}
\affiliation{Department of Physics \& Astronomy, University of New Mexico, 1919 Lomas Boulevard NE, Albuquerque, NM 87131, USA}

\author[0000-0001-8014-0270]{Patrick J. Lowrance}
\affiliation{IPAC-Spitzer, California Institute of Technology, MC 314-6, 1200 E. California Blvd., Pasadena, California 91125, USA}

\author[0000-0003-4150-841X]{Elisabeth R. Newton}
\affiliation{Department of Physics and Astronomy, Dartmouth College, Hanover NH 03755, USA}


\begin{abstract}

We present the observation of a secondary eclipse of the young hot Neptune, AU Mic b, in the infrared using the Spitzer Space Telescope. Using a primary transit from Spitzer to constrain the system parameters, we tentatively detect an eclipse centered at BJD=$2458740.848893^{+0.00010}_{-0.000099}$ with an observed depth of 171$\pm{29}$ ppm given an uninformed prior. This corresponds to a dayside brightness temperature of T=1031$\pm{58}$ K, which exceeds the calculated equilibrium temperature of 606$\pm{19}$ K. We explore some possible explanations for these results, including inefficient heat redistribution, gravitational contraction, stellar pulsations, instrument systematics and choice of eclipse depth prior, but find none of these to be likely explanations for the observed eclipse parameters. We also explore the impact of correlated noise in the systematic trends, and we find that splitting the systematics into low-pass (smoothing) and high-pass trends is required to reach an optimal minimization of the low-frequency systematics in the resulting detrended light curve. Future observations with JWST are needed to confirm our eclipse detection with Spitzer. \\

\end{abstract}


\section{Introduction} \label{sec:intro}

More than 6,000 validated and/or confirmed exoplanets have been discovered to date, with nearly 75\% of those transiting in front of their host stars \citep{akeson2013}\footnote{\url{https://exoplanetarchive.ipac.caltech.edu}}. Transiting exoplanets provide a unique window into exoplanet properties such as directly measuring the planetary radius. When combined with radial velocities or transit timing variations, the bulk densities of transiting exoplanets can be measured \citep[e.g.,][]{chenkipping2017,zeng2019}, and through transmission spectroscopy the presence of clouds, hazes and different atomic and molecular atmospheric abundances inferred \citep[e.g.,][]{Charbonneau2002, Vidal-Madjar2004, Redfield2008, Sing2011, Birkby2013, gao2018}. In addition, observations of secondary eclipses, when an exoplanet passes behind its host star, offer the opportunity to directly measure the exoplanet's thermal emission by giving a measure of the flux coming directly from the dayside of the planet, as well as phase curves at other illumination angles \citep[e.g.,][]{knutson2008,shporer2011}. Secondary eclipses of many Jovian exoplanets have previously been detected -- e.g., HD 209458 b \citep{deming2005}, TrES-1 b \citep{charbonneau2005}, HD 189733 b \citep{deming2006}, and XO-1 b \citep{machalek2008}. Detecting thermal eclipses is challenging because at visible wavelengths the flux contrast is not favorable, with hot Jupiters producing eclipse depths of 100 ppm or less, requiring space-based observatories to detect for individual systems \citep{Rowe2008, Christiansen2010, Coughlin2012, Morris2013, Angerhausen2015, Mansfield2018, Lendl2020} and in the ensemble \citep{Sheets2017}. Further, there is limited observing time on mid-infrared space-based observatories such as the James Webb Space Telescope (hereafter \textit{JWST}), and the now decommissioned Spitzer Space Telescope (hereafter \textit{Spitzer}), where the flux contrast is greatest. Detecting thermal eclipses is also limited by photon statistics, due to the relative flux contrast. Therefore, the majority of planets with thermal emission measurements are the hottest of the hot Jupiters orbiting the brightest nearby host stars.

In particular, there are only a small number of secondary eclipse detections to date of Neptune or terrestrial exoplanets: 55 Cnc e \citep{demory2012}, GJ 436 b \citep{deming2007, Gillon2007}, GJ 3470 b \citep{benneke2019}, TRAPPIST-1 b \citep{Greene2023}, GJ 1132 b \citep{Xue2024}, LTT 1445A b \citep{Wachiraphan2025}, and TOI-1468 b \citep{MeierValdes2025}. Furthermore, there are no published secondary eclipse detections of young exoplanets with ages $<$100 Myr. The number of known young exoplanets transiting GKM dwarfs has increased significantly with the \textit{K2} \citep{howell2014} and \tess \citep{ricker2015} missions. These young systems can inform our knowledge of planet formation by providing single-system snapshots of the dynamical evolution of planetary systems, rather than inferring the dynamical evolution from the demographics of more mature planetary systems \citep[e.g.,][]{Dressing2013, Pascucci2019, Mulders2020, Mulders2021, Fernandes2019, Howard2012}. Measuring the thermal emission of young exoplanets can help distinguish between hot-start and cold-start models for planet formation, and constrain theoretical simulations of cooling curves \citep{Linder2019,Mordasini2012,burrows1997,Hansen2012}.

AU Microscopii (\aumic) is the second-closest \citep[9.7 pc,][]{bailer-jones2018} pre-main sequence star to the Sun \citep[$22 \pm 3$ Myr,][]{mamajek2014}. We summarize the stellar properties in Table~\ref{table:stellarquant}. \aumic is a member of the Beta Pictoris Moving Group, which also hosts the planets Beta Pic b and c confirmed through radial velocities, direct imaging, astrometric motion and interferometry \citep{lagrange2010, lacour2021, vandal2020, nowak2020, snellen2018}. \aumic is the host to a pair of young Neptune-sized transiting exoplanets, AU Mic b \citep{plavchan2020} and AU Mic c \citep{Gilbert2022, martioli2021}, as well as an Earth-mass planet validated through transit-timing variations (TTVs), AU Mic d \citep{Wittrock2023, Boldog2025}, which all together comprise a 4:6:9 resonant chain. We summarize the planetary properties for all three planets in Table~\ref{table:planetquant}. Given their youth, these systems are crucial probes of the planet formation process. \aumic b is spin-orbit aligned with the spin of its host star, and is in a stellar rotation to orbital period 7:4 resonance \citep{addison2021, Szabo2021, palle2020, martioli2020, Hirano2020}. Somewhat surprisingly, \citet{Cale2021,Zicher2022,Baptiste2022} have found evidence from RVs that the planets may be denser than expected, resembling the density of Neptunes orbiting much older stars. A similar result was also recently found for two of the Jovian planets in the similar age v1298 Tau resonant chain of planets \citep{SuarezMascareno2021, Maggio2022}.

\aumic is also thought to possess a strong stellar wind, which in addition to UV-driven photo-evaporation \citep{Ribas2005, Sanz-Forcada2011,Pascucci2019}, may strongly impact the evolution of the atmospheres of its young exoplanets \citep{Strubbe2006,plavchan2005,plavchan2009}. \citet{carolan2020} and \citet{Alvarado-Gomez2022} investigate the atmospheric confinement caused by stellar winds for \aumic b. Another way to probe the atmospheric status and evolution of the planets of the \aumic system are by looking for phase curves, transmission spectroscopy, and/or thermal eclipses. No phase curves of the \aumic planets have been attempted to date due to the large amount of stellar variability. A search for atmospheric helium emission during the transit of \aumic b was inconclusive \citep{martioli2020}, and no atmospheric sodium emission was observed by ESPRESSO during another transit of \aumic b \citep{palle2020}.

In this work, we present measurements from a search for a thermal eclipse of AU Mic b with the \textit{Spitzer} Space Telescope. This paper is organized as follows: In ${\S}$\ref{sec:observations}, we present the observations of AU Mic b from \spitzer\, the extraction of the primary and secondary light curves, and basic data reduction and photometric analysis, as well as supplementary light curves. In ${\S}$\ref{sec:analysis}, we describe the treatment of stellar activity, light curve detrending, and the process of modeling the data with the \exofast package \citep{eastman2019}. ${\S}$\ref{sec:results} reports the derived posterior distributions of the analysis, including the secondary eclipse depth. An interpretation of the results follows in ${\S}$\ref{sec:discussion}, including calculations of the dayside temperature, a discussion of AU Mic b formation evolution simulations, an exploration of possible explanations for the observed results, and a thorough review of the detrending methodology used. Finally, ${\S}$\ref{sec:conclusion} provides a summary of our findings.

\begin{deluxetable}{l|c|c|c}
    \tablecaption{\label{table:stellarquant}Stellar properties for host star \aumic.}
    \tablehead{Property & Unit & Quantity &  Ref}
    \startdata
Spectral Type & ...       & M1Ve                  & 1   \\
\magv         & mag       & 8.81 $\pm$ 0.10       & 1   \\
\magt         & mag       & 6.755 $\pm$ 0.032     & 1   \\
\alphaj       & h:m:s     & 20:45:09.53           & 2   \\
\deltaj       & deg:am:as & -31:20:27.24          & 2   \\
\pmalpha      & mas/yr    & 281.424 $\pm$ 0.075   & 2   \\
\pmdelta      & mas/yr    & -359.895 $\pm$ 0.054  & 2   \\
Distance      & pc        & 9.7221 $\pm$ 0.0046   & 3   \\
Parallax      & mas       & 102.8295 $\pm$ 0.0486 & 2   \\
\mstar        & \msun     & 0.50 $\pm$ 0.03       & 4   \\
\rstar        & \rsun     & 0.862 $\pm$ 0.052     & 5   \\
\teff         & K         & 3\,700 $\pm$ 100      & 6   \\
\lstar        & \lsun     & 0.09                  & 6   \\
Age           & Myr       & 22 $\pm$ 3            & 7   \\
\prot         & days      & 4.8367 $\pm$ 0.0006   & 8  \\
v $\sin{i}$   & km $s^{-1}$      & 8.7 $\pm$ 0.2         & 9  \\
    \enddata
    \tablerefs{(1) \citet{Stassun2019}; (2) \citet{gaia2018}; (3) \citet{bailer-jones2018}; (4) \citet{plavchan2020}; (5) \citep{Gallenne2022}; (6) \citet{plavchan2009}; (7) \citet{mamajek2014}; (8) \citet{Szabo2021}; (9) \citet{lannier2017}}
\end{deluxetable}

\begin{deluxetable*}{l|c|c|c|c|c|c}
    \tablecaption{\label{table:planetquant}Planetary properties for \aumic system.}
    \tablehead{Property & Description & Unit & \aumic b & \aumic c & \aumic d & Ref}
    \startdata
\porb                         & Orbital Period                    & days       & 8.4630004$^{+0.0000058}_{-0.0000060}$ & 18.858982$^{+0.000053}_{-0.000050}$ & 12.73596 $\pm$ 0.00793 & 1, 2 (d) \\
a                             & Semi-Major Axis                   & au         & 0.0645 $\pm$ 0.0013                   & 0.1101 $\pm$ 0.0022         & 0.0853 $\pm$ 0.0016 & 3, 2 (d) \\
e                             & Eccentricity                      & ...        & 0.00577 $\pm$ 0.00101                      & 0.00338 $\pm$ 0.00164  & 0.00305 $\pm$ 0.00104 & 2 \\
i                             & Inclination                       & deg        & 89.18 $^{+0.53}_{-0.45}$              & 89.39 $^{+0.40}_{-0.38}$                & 89.31812$\pm$ 1.15800 & 1, 2 (d) \\
$\omega$                      & Argument of Periastron            & deg        & 88.43038 $\pm$ 0.05783               & 223.28438 $\pm$ 1.68357                & 160.78945 $\pm$ 2.59947 & 2 \\ 
\multirow{2}{*}{\mplan}       & \multirow{2}{*}{Planetary Mass}   & \mjup      & 0.063 $\pm$ 0.005                     & 0.030 $\pm$ 0.007                   & 0.00331 $\pm$ 0.00161 & \multirow{2}{*}{4, 2 (d)} \\
                              &                                   & \mear      & 20.12$^{+1.57}_{-1.72}$               & 9.60$^{+2.07}_{-2.31}$              & 1.053 $\pm$ 0.511 &   \\
\multirow{2}{*}{\rplan}       & \multirow{2}{*}{Planetary Radius} & \rjup      & 0.374$^{+0.021}_{-0.020}$             & 0.249$^{+0.028}_{-0.027}$           & ... &  \multirow{2}{*}{1} \\
                              &                                   & \rear      & 4.19$^{+0.24}_{-0.22}$                & 2.79$^{+0.31}_{-0.30}$              & ... &   \\
$\rho_{\mbox{\scriptsize p}}$ & Planetary Density                 & g/cm$^{3}$ & 1.32$^{+0.19}_{-0.20}$                & 1.22$^{+0.26}_{-0.29}$              & ... & 4 \\
K                             & RV Semi-Amplitude                 & m $s^{-1}$        & 10.23$^{+0.88}_{-0.91}$               & 3.68$^{+0.87}_{-0.86}$              & 0.45150 $\pm$ 0.21874 & 4, 2 (d) \\
\tc $-$ 2\,458\,000           & Time of Conjunction               & BJD        & 330.39051 $\pm$ 0.00015               & 342.2223 $\pm$ 0.0005               & 340.55781 $\pm$ 0.11641 & 3, 2 (d) \\
\tdur                         & Transit Duration                  & hours      & 3.50 $\pm$ 0.08                       & 4.5 $\pm$ 0.8                       & ... & 3 \\
\rplan/\rstar                 & Transit Depth                     & ...        & 0.0512 $\pm$ 0.0020                   & 0.0340$^{+0.0034}_{-0.0033}$        & ... & 1 \\
a/\rstar                      & ...                               & ...        & 19.1 $\pm$ 0.3                        & 29 $\pm$ 3                          & ... & 3 \\
b                             & Impact Parameter                  & ...        & 0.16$^{+0.13}_{-0.11}$                & 0.30$^{+0.21}_{-0.20}$              & ... & 1
    \enddata
    \tablerefs{(1) \citet{Gilbert2022}; (2) \citet{Wittrock2023}; (3) \citet{martioli2021}; (4) \citet{Cale2021}}
\end{deluxetable*}

\section{Observations and Light Curve Extraction} \label{sec:observations}

In ${\S}$\ref{sect:spitzerobs}, we present the primary observations gathered for this paper from the \textit{Spitzer} Space Telescope, and in ${\S}$\ref{sect:spitzerinitdetrend} we present the initial \textit{Spitzer} light curve detrending with instrumental systematics. In ${\S}$\ref{sect:tessobs}, we present the \textit{TESS} light curve data used to estimate the stellar variability from the rotational modulation of active regions at the non-contemporaneous times of the \textit{Spitzer} observations. In ${\S}$\ref{sect:evryscope}, we present observations of AU Mic collected with Evryscope.

\subsection{\textit{Spitzer}/IRAC Photometry and Light Curve Extraction}
\label{sect:spitzerobs}
\spitzer\ Director’s Discretionary Time observations were awarded (DDT; PID 14214, 14241) for observations of AU Mic with the Infrared Array Camera (\irac: \citealt{fazio2004}) to collect validation data on the planetary candidate detected by \tess. Three observations at the predicted time of transit of AU Mic b were obtained, along with an observation at an incorrectly predicted transit time of AU Mic c, and a predicted time for a secondary transit of AU Mic b. The first primary transit of AU Mic b is presented in \citet{plavchan2020}, and all three in \citet{Wittrock2022}. In this work, our analysis focuses on the third primary transit of AU Mic b from 2019 September 10 (BJD=2458736), and the immediately following secondary eclipse on 2019 September 14 (BJD=2458740).

Both of the observations were taken using the 32x32 pixel sub-array mode with an exposure time of 0.08 s to avoid saturation on the star (measurement cadence is 0.1\,s). After placing the star on the ``sweet-spot'' pixel, using the pointing calibration and reference sensor (PCRS) peak-up mode \citep{ingalls2012}, we exposed with continuous staring (no dithers). The observations were all taken at at 4.5 $\mu$m, as this channel has lower systematics due to the intra-pixel sensitivity. The coordinates were adjusted for the high parallax and proper motion of AU Mic for the proposed observation dates. Each observation set consisted of a 30-min pre-stare dither pattern, an 8 hour stare, and a 10 minute post-staring dither pattern. All data were calibrated by the \spitzer\ pipeline S19.2 and can be accessed using the \spitzer\ Heritage Archive (SHA)\footnote{\url{http://sha.ipac.caltech.edu}}.

We then follow the same data reduction and light curve extraction procedure as in \citet{demory2012} and  \citet{Wittrock2022}, which we briefly summarize herein. After converting fluxes from specific intensity to photon counts, we perform aperture photometry on each subarray image using the IDL Astronomy User's Library\footnote{\url{http://idlastro.gsfc.nasa.gov/contents.html}} routine \texttt{APER}. We compute stellar fluxes using aperture radii between 2.0 and 4.0 pixels in increments of 0.2 pixels in aperture radius, and subtracting a sky annulus extending from 11 to 15.5 pixels from the center of the point response function (PRF). The PRF center and full width at half maximum (FWHM) along both x and y detector pixel axes is measured by fitting a Gaussian profile on each image using the MPCURVEFIT procedure \citep{markwardt2009}. For each block of 64 sequential subarray images in time, we discard discrepant values for the measurements of flux, background (noise/pixel), x-y positions and FWHM (in x and y) using a 10-$\sigma$ median clipping for these six parameters. We then average the resulting values across the remaining subset of the 64 subarray images, with the photometric flux rms being taken as the uncertainties on the average flux measurements. At this stage, a 50-$\sigma$ clipping moving average is used on the resulting light curve to remove very discrepant subarray-averaged fluxes. This process is repeated for every aperture radius considered. For both the primary transit and secondary eclipse observations, the ``best'' results are obtained using a 3.2 pixel aperture radius, where we quantify the ``best'' as the extracted light curve with the smallest out of transit RMS and thus maximum SNR and minimal systematics.

\begin{figure*}
    \centering
    \begin{tabular}{cc}
    \includegraphics[width=80mm]{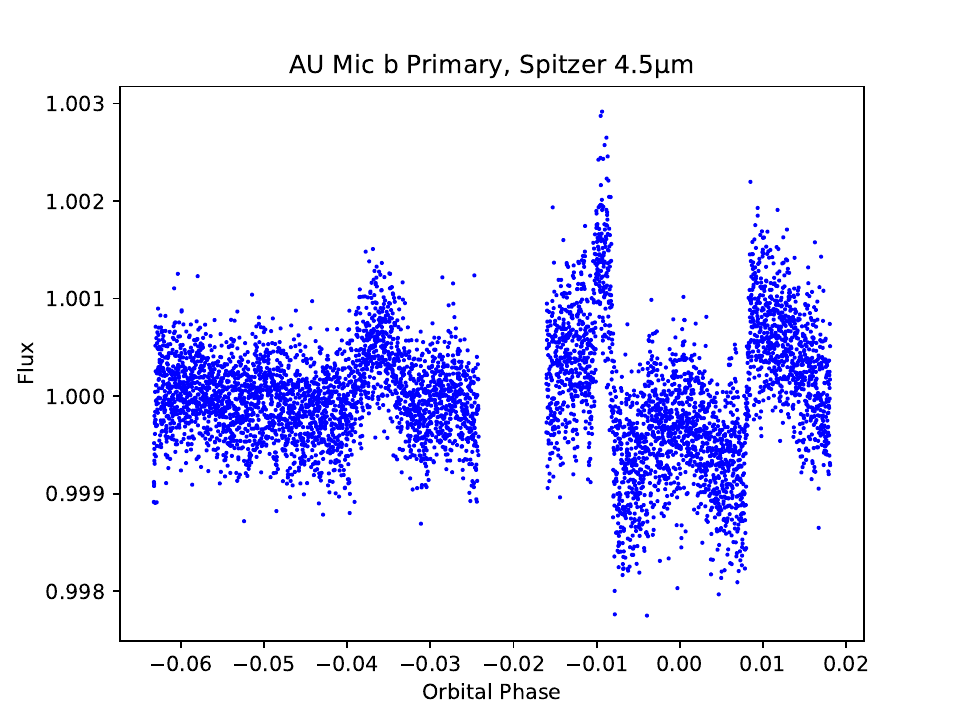} &
    \includegraphics[width=80mm]{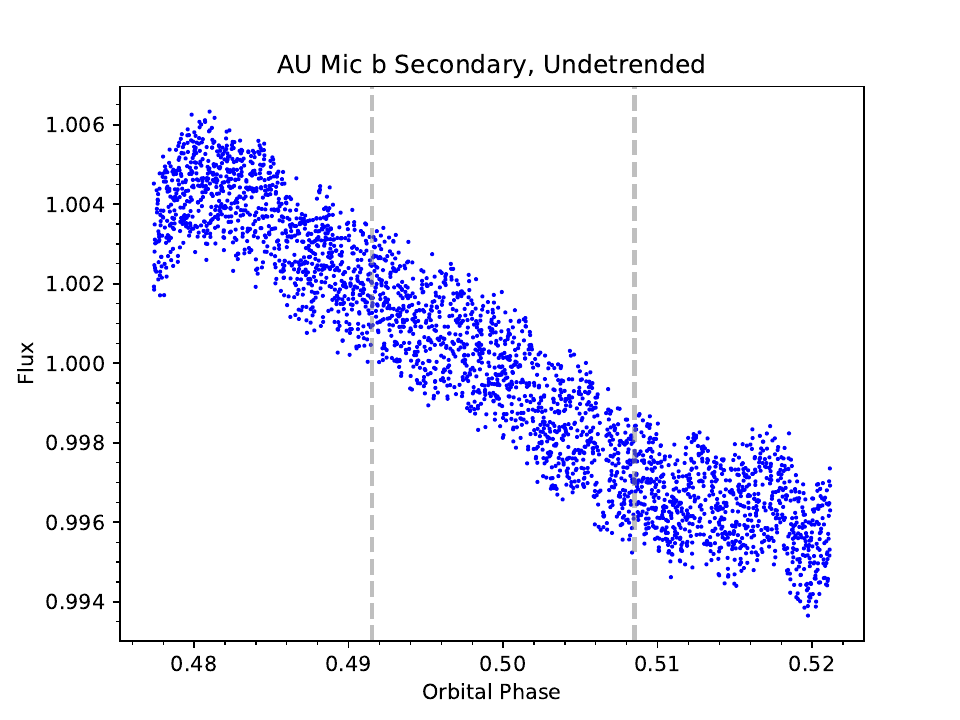} \\
    \end{tabular}
    \caption{\textit{Left}: The detrended and phased \spitzer/\irac 4.5 $\mu$m photometry of the primary transit of AU Mic b. \textit{Right}: The undetrended and phased \spitzer/\irac 4.5 $\mu$m secondary eclipse light curve. The vertical dashed lines correspond to the predicted time of eclipse ingress and egress for a circular orbit.}
    \label{fig:spitzerLCs}
\end{figure*}

\subsection{\textit{TESS} Light Curve}
\label{sect:tessobs}
\tess\ observed \aumic in Cycle 1, Sector 1 from 2018 July 25 to 2018 August 22 at a 2-minute cadence \citep{plavchan2020} and again in Cycle 3, Sector 27 from 2020 July 05 to 2020 July 30 at a 20-second cadence, with 2-minute cadence data also generated by co-adding sets of six 20-second exposures \citep{Gilbert2022}.

\subsection{EvryScope Light Curve}
\label{sect:evryscope}

\begin{figure*}
    \centering
    \begin{tabular}{cc}
    \includegraphics[width=80mm]{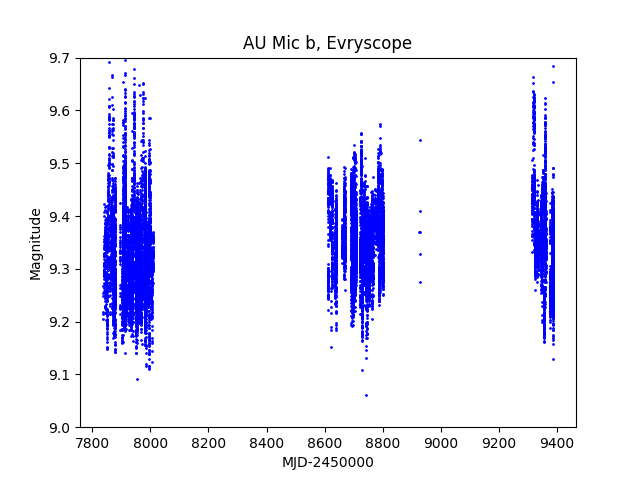} & 
    \includegraphics[width=80mm]{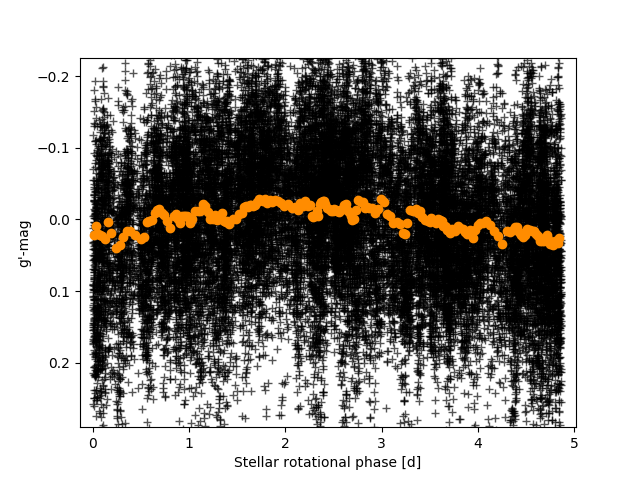} \\
    \end{tabular}
    \caption{Photometry of Evryscope observations  \textit{Left}: The full Evryscope light curve. \textit{Right}: Phase-folded light curve of Evryscope observations. Data in black, binned data in orange.}
    \label{fig:evryscope}
\end{figure*}

EvryScope observed the southern sky using an array of small cameras on a custom mount \citep{Howard2019, Howard2020, Ratzloff2019}. Approximately 27,000 epochs of AU Mic were collected at a two-minute cadence in the $g^\prime$ filter in the 2017-2019 seasons, crucially sampling the light curve inbetween the 2018 and 2020 27-day \tess\ sector observations. The data collected at a two-minute cadence was not continuous for whole nights; this cadence for AU Mic was typically achieved for approximately a one hour baseline per observable night (Figure \ref{fig:evryscope}). At $g^\prime = 9.5$, AU Mic is close to the non-linear saturated regime for Evryscope, resulting in a single measurement precision of $\sim$10\%. Phase-folding the light curve recovers the stellar rotation period (Figure \ref{fig:evryscope}), demonstrating that the spot modulation pattern is stable to the precision of Evryscope over the 2017-2019 seasons, which span and include the first \tess\ observations and the \spitzer\ observations. We also bin the EvryScope light curve into per-hour bins to reduce the shot noise for comparison to the rotational modulation photometric variability, which is smaller than the shot noise amplitude of $\sim$10\%.

\section{Data Analysis} \label{sec:analysis}

Here we present the modeling used for the transit and eclipse light curves. First in ${\S}$\ref{sect:analysis_activity}, we consider the impact of stellar variability on the \spitzer\ light curve. In ${\S}$\ref{sect:spitzerinitdetrend} we present the initial detrending of the \spitzer primary transit light curve, and in ${\S}$\ref{sect:analysis_parametric} we then describe the parametric model used to account for the long-term -- relative to the \spitzer\ time baseline -- stellar activity trends present in both \spitzer\ light curves. Finally in ${\S}$\ref{sect:analysis_exofast}, we provide the final joint modeling of both the primary transit and secondary eclipse performed with \exofast \citep{eastman2019}.

\subsection{Characterization of rotational modulation of the stellar activity during the \spitzer\ observations from the \tess\ observations}
\label{sect:analysis_activity}

There is a fortuitous 7:4 stellar spin -- planetary orbit period commensurability between the rotation of the host star AU Mic and the orbital period of AU Mic b. This was first discovered in \citet{Szabo2021}. This implies that every fourth transit of AU Mic b transits the same or nearly the same range of stellar longitudes. Additionally, AU Mic possessed a long-lived spot pattern that spanned multiple years as seen by \tess\ in 2018 and 2020 \citep[Figure 8, ][]{Gilbert2022}, which also fortuitously bound the \spitzer\ observations in 2019. This is best exemplified by the nearly identical positioning of the AU Mic b transits in \tess\ with respect to the out-of-transit photometric variability from the rotational modulation of the active regions -- e.g. Figure 9 in \citet{Gilbert2022} and Figure 1 in \citet{martioli2021}. Consequently, we can use the long-lived active regions of AU Mic and the period commensurability to predict the photometric variability due to stellar activity during the intervening two years, including when our \spitzer\ observations were obtained, from the \tess\ light curves\footnote{As a first caveat to the assumption of period commensurability made herein, we cannot rule out the possibility that there is a $(N\pm 1)/N$ error in the stellar rotation period as discussed in \citet{Szabo2021}, where $N$ is the number of stellar rotation cycles between \tess\ Cycles 1 and 27. However, the agreement of this period commensurability to 0.1\% from the \tess\ light curves alone is compelling. Second, we cannot rule out the possibility that AU Mic took on a completely different spot modulation pattern of photometric variability in 2019 before returning in 2020 to the nearly identical rotational variability pattern seen by \tess\ in 2018 at this period commensurability. There does not exist sufficient ancillary light curve information that was collected on the ground or in space in 2019 to definitively rule these caveats out. We do not further consider these caveats herein, and future observations beyond the scope of this work will enable further assessment of period commensurability and active region longevity. This remains a limiting assumption of the analysis presented herein}.

\begin{figure*}
    \centering
    \begin{tabular}{ccc}
    \includegraphics[width=80mm]{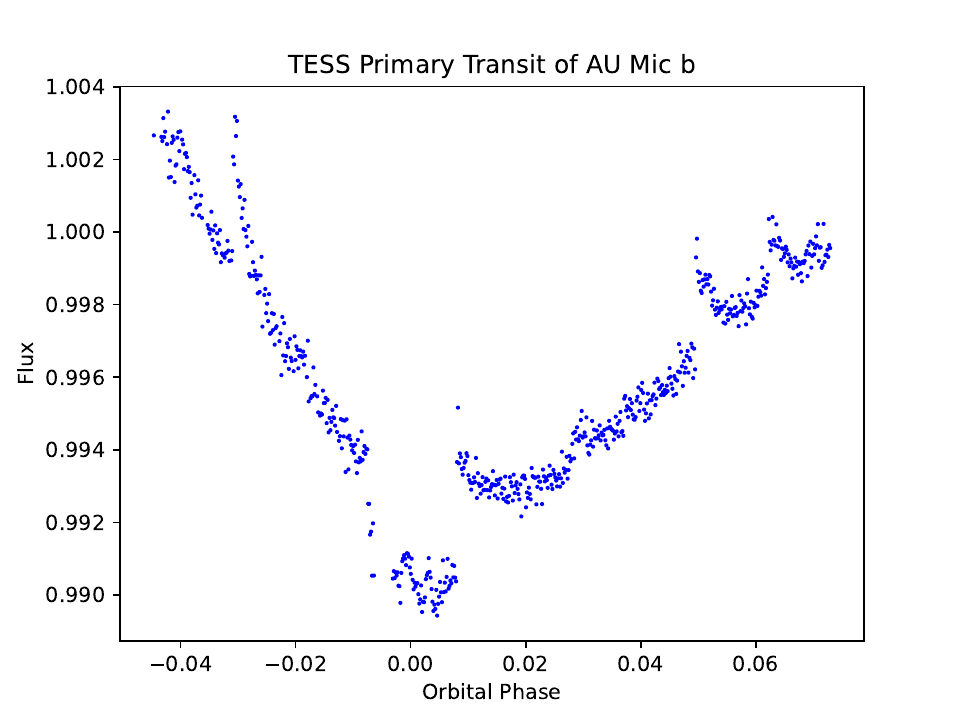} & \includegraphics[width=80mm]{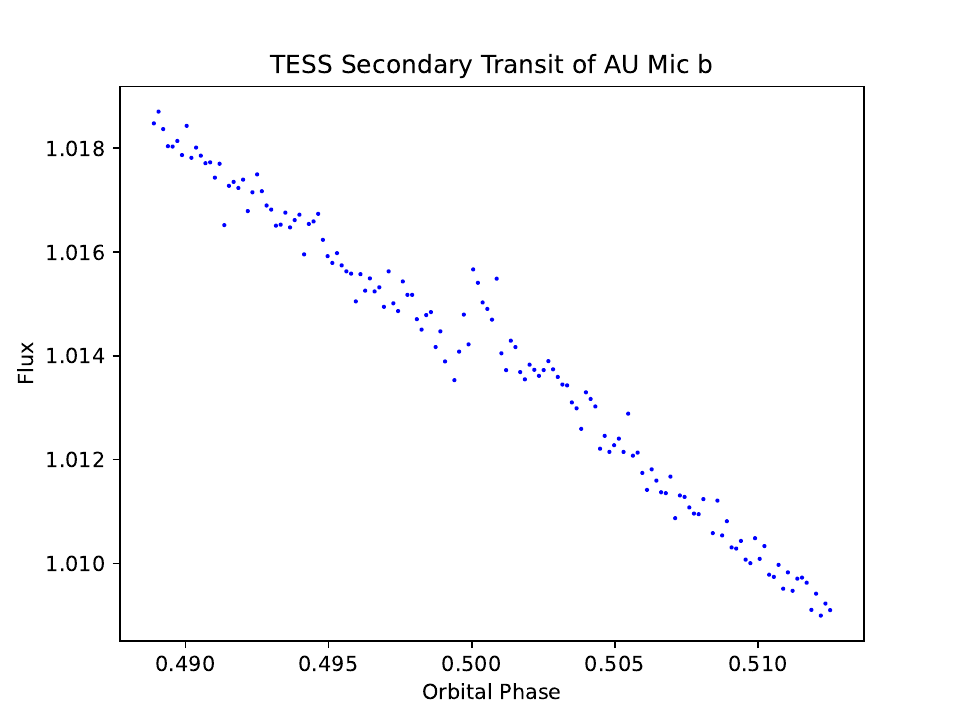} \\
    \end{tabular}
    \caption{\tess\ photometry of AU Mic b at the same stellar activity phase as the \spitzer\ observations assuming the spin-orbit commensurability and activity stability between \tess cycles 1 and 3. \textit{Left}: A portion of the Sector 1 \tess\ light curve around the first \tess\ transit of \aumic b, epoch 0. \textit{Right}: The \tess\ light curve at epoch 0.5, around the time of predicted secondary eclipse. In the \tess light curve on the right a complex flare morphology is visible on top of the decreasing linear flux trend at the predicted time of eclipse. See \citet{Gilbert2022} for flare analysis.}
    \label{fig:tess}
\end{figure*}

The first \tess\ transit of AU Mic b is at BJD=2458736 as epoch 0. The primary transit observed with \spitzer occurred at transit epoch 48, meaning it is at the same stellar longitude of the stellar activity (as 48\%4=0). Consequently, from the TESS light curve at epoch 0, we can forecast what the photometric variability at \tess\ wavelengths should have been doing at the time of the first \spitzer\ primary observation. As seen in the \tess\ light curve of this transit, the rotational modulation of stellar activity results in a photometric variability signal that resembles a ``concave-up'' quadratic (positive coefficient in the second-order term) with the transit occurring just prior to the minimum of the out of transit photometric variability (Figure \ref{fig:tess}).

However, comparing the two light curves shows that the highly variable parabolic activity trend in the \tess\ data is not present in the detrended \spitzer\ transit, illustrating that the effects of the stellar activity are significantly diminished at the \spitzer\ wavelength of 4.5 $\mu$m. Furthermore, we examine the \tess\ data around the next predicted time of secondary eclipse, epoch 0.5, and find only a constant negative linear trend, indicating that any thermal eclipse signal we detect in the \spitzer data is likely not caused by a signal from the stellar activity of AU Mic. Figure~\ref{fig:tess} shows the \tess\ light curve both at epochs 0 and 0.5.

\subsection{Initial \textit{Spitzer}/IRAC Primary Transit Light Curve Detrending}
\label{sect:spitzerinitdetrend}
All IRAC photometry at 4.5 $\mu$m contains instrumental systematics caused by the coupling between spacecraft pointing fluctuations, spacecraft heater-induced temperature changes, and drifts with intra-pixel sensitivity variations. Therefore, we use the initial detrending of the primary light curve from \citet{Wittrock2022} to detrend the instrument systematics using both the trend time series for the pixel centroid motion and the PSF FWHM, which is summarized herein.

We model the IRAC intra-pixel sensitivity (\citep{ingalls2016}) using a modified implementation of the \texttt{BLISS} (\texttt{BiLinearly-Interpolated Sub-pixel Sensitivity}) mapping (BM) algorithm \citep{stevenson2012}, complementing the BM correction with a linear function of the point response function FWHM. In addition to the BM, our baseline model includes the PRF’s FWHM along the x and y axes, which significantly reduces the level of correlated noise as shown in previous studies (e.g., \citep{Lanotte2014}, \citep{Demory2016a}, \citep{Demory2016b}, \citep{Gillon2017}, \citep{mendonca2018}).

\subsection{Parametric models for the primary and secondary observations non-transit-like photometric variability}
\label{sect:analysis_parametric}
As can be seen from the primary transit light curve, there is significant non-transit like photometric variability present in the primary transit observation. First, we see an unusual brightening ``bump'' prior to the primary transit, as well as an unusual briefer pre-ingress brightening. Third, we see by eye a broad brightening that peaks at mid-transit as might be expected from a limb-brightening effect, but this also extends both before ingress and after egress. None of these three features correlate with any of the expected \spitzer\ systematic trend time-series with centroid-motion, thermal heating cycles or FWHM, and are therefore likely astrophysical in origin. These unique photometric variability features are to be discussed further in future work, but their astrophysical origins, such as potential photospheric stellar activity, are not directly considered herein. As in \citet{Wittrock2022}, we use the same 2nd-order polynomial model fit to account for this stellar activity behavior, as well as the same ad hoc Gaussian model given the as-of-yet unexplained presence of a low-level Gaussian-like trend coincident with the transit midpoint time. 

Similarly, we generate a 2nd-order polynomial model for the stellar activity in the secondary transit light curve. We further identify a systematic deviation in multiple systematic time-series that is not entirely masked with the sigma-clipping in ${\S}$\ref{sect:spitzerobs}. This in turn can produce an apparent detrended flux deviation resembling a low amplitude flare (low with respect to the flux RMS). Given the proximity of this systematic to the egress time, we do not remove these data. This systematic jump occurs mid-eclipse from BJD\_TDB 2458740.90 to 2458740.91.

\subsection{\exofast modeling of the \spitzer\ light curves}
\label{sect:analysis_exofast}
\begin{deluxetable}{l|c|c|c}
    \tablecaption{\label{table:priors}Initial stellar and planetary priors for \exofast modeling of the \aumic secondary eclipse.}
    \tablehead{Prior & Unit & Prior Center & Width}
    \startdata
\logmstmsu      & ...       & -0.3        & 0.026   \\
\rstar          & \rsun     & 0.862       & 0.052   \\
\teff           & K         & 3\,700      & 100   \\
$[{\rm Fe/H}]$  & ...       & 0           & 0   \\
Age             & Myr       & 22          & 3   \\
\porb           & days       & 8.46321 & 0.00004 \\
\rplan/\rstar   & ... & 0.0514 & 0.0013 \\
\tc             & BJD        & 2458736.61729 & 0.00014  \\
Limb darkening coeff. u1 & ...  & 0.17 &  0.22 \\
Limb darkening coeff. u2 & ...  & 0.15 & 0.27 \\
$\cos(i)$       & deg       & 0.0084    & 0.0074 \\
\enddata
\end{deluxetable}

We use the \exofast package \citep{eastman2019} to jointly model the primary and secondary light curves and characterize the thermal eclipse by estimating the posterior probabilities through the Markov Chain Monte Carlo (MCMC) method to determine the statistical significance of our thermal eclipse detection. We apply a low-pass Savitzky-Golay filter (smoothing width of 15 minutes) to the systematic time-series, but not the flux data. We then subtract the smoothed trends from the original systematic trends to generate high-pass filtered systematic time-series. This particular timescale was chosen to be shorter than the known \spitzer systematic effects from the on-board heater cycle of one hour, and longer than the expected eclipse ingress and egress times of approximately 10 minutes. The justification for the filtering/smoothing of the systematic time-series is discussed in more detail in ${\S}$\ref{sect:detrendingpalooza}. We provide to \exofast all of the smoothed and filtered detrending parameters (sky brightness, noise/pixel, x-y positions, and FWHM in x and y), as well as the parametric models discussed in ${\S}$\ref{sect:analysis_parametric}, as multiplicative detrending parameters. 

The initial stellar and planetary Gaussian priors were taken from \citep{plavchan2020}, with the exception of an updated \rstar value from \citep{Gallenne2022}, and are listed in Table~\ref{table:priors}. We also enforce a circular orbit, as eccentric models are more computationally intensive and preliminary runs found eccentricities that were either unphysical or consistent with a circular orbit but without providing a more robust fit. We then perform the MCMC run, integrating over 250,000 steps or until convergence is reached, where convergence is defined as achieving a Gelman-Rubin statistic less than 1.01 and an effective number of independent draws $T(z)$ greater than 1000 \citep{Ford2006, eastman2019}. Upon completion, \exofast generates the posterior probability distributions, median parameter values, and primary and secondary transit models.

\section{Results} \label{sec:results}

The median values and 68\% confidence intervals generated by \exofast for the fitted stellar and planetary parameters are given in Tables~\ref{table:exfstellar} and \ref{table:exfplanet}, respectively. Figure \ref{fig:binnedmodel} shows the best fit model of the secondary eclipse light curve. The \exofast generated MCMC corner plot and posterior probability distribution functions are provided in the Appendix. The key result from this analysis is the statistically robust detection of thermal emission from the planet, A$_{\mbox{\scriptsize T}}$, of 171$\pm{29}$ ppm. The thermal emission detected from \aumic b corresponds to a brightness temperature of 1031$\pm{58}$ K, much higher than the \exofast fitted equilibrium temperature of 606$\pm{19}$ K (assuming no albedo and perfect heat redistribution). The implications of this discrepancy are discussed further in ${\S}$\ref{sec:discussion}.

\begin{deluxetable*}{l|c|c|c}[ht]
\tablecaption{\label{table:exfstellar}\exofast-generated median values and 68\% confidence interval for host star \aumic.}
\tablehead{Posterior & Description & Unit & Quantity}
\startdata
\mstar             & Stellar Mass                     & \msun      & $0.520^{+0.032}_{-0.030}$ \\
\rstar             & Stellar Radius                   & \rsun      & $0.754^{+0.031}_{-0.025}$ \\
\lstar             & Stellar Luminosity               & \lsun      & $0.097^{+0.014}_{-0.012}$ \\
$\rho_{\star}$     & Stellar Density                  & g/cm$^{3}$ & $1.72^{+0.16}_{-0.18}$ \\
$\log{g}$          & Surface Gravity                  & ...        & $4.400^{+0.027}_{-0.034}$ \\
\teff              & Effective Temperature            & K          & $3700\pm100$         \\
Age                & ...                              & Gyr        & $0.0221\pm0.003$ 
\enddata
    \tablecomments{See Table 3 in \citet{eastman2019} for a detailed description of all parameters.}
\end{deluxetable*}

\startlongtable
\begin{deluxetable*}{l|c|c|c}
    \tablecaption{\label{table:exfplanet}\exofast-generated median values and 68\% confidence interval for planet AU Mic b.}
    \tablehead{Posterior & Description & Unit & Quantity}

\startdata
\porb                                                 & Orbital Period                               & days                      & $8.463210\pm0.000041$      \\
\mplan                                                & Planetary Mass\tbnm{a}                       & \mjup                     & $0.053^{+0.020}_{-0.013}$        \\
\rplan                                                & Planetary Radius                             & \rjup                     & $0.352^{+0.016}_{-0.012}$         \\
\tc                                                   & Time of Conjunction\tbnm{b}                  & BJD\_TDB                  & $2458736.617288\pm0.000097$      \\
T$_{\mbox{\scriptsize T}}$                            & Time of Minimum Projected Separation\tbnm{c} & BJD\_TDB                  & $2458736.617288\pm0.000097$      \\
T$_{\mbox{\scriptsize 0}}$                            & Optimal Conjunction Time\tbnm{d}             & BJD\_TDB                  & $2458736.617288\pm0.000097$    \\
a                                                     & Semi-Major Axis                              & au                        & $0.0654\pm0.0013$               \\
i                                                     & Inclination                                  & deg                       & $89.12^{+0.41}_{-0.35}$       \\
T$_{\mbox{\scriptsize eq}}$                           & Equilibrium Temperature\tbnm{e}              & K                         & $606\pm19$                \\
$\tau_{\rm circ}$                                     & Tidal Circularization Timescale              & Gyr                       & $199^{+77}_{-50}$                  \\
K                                                     & RV Semi-Amplitude\tbnm{a}                    & m/s                       & $8.2^{+3.1}_{-1.9}$               \\
\rplan/\rstar                                         & ...                                          & ...                       & $0.04801\pm0.00042$       \\
a/\rstar                                              & ...                                          & ...                       & $18.66^{+0.56}_{-0.69}$             \\
$\delta$                                              & Transit Depth                                & ...                       & $0.002305^{+0.000041}_{-0.000040}$      \\
Depth                                                 & Flux Decrement at Mid Transit                & ...                       & $0.002305^{+0.000041}_{-0.000040}$      \\
$\tau$                                                & Ingress/Egress Transit Duration              & days                      & $0.00724^{+0.00060}_{-0.00044}$     \\
T$_{14}$                                              & Total Transit Duration                       & days                      & $0.14561^{+0.00060}_{-0.00053}$      \\
T$_{\mbox{\scriptsize FWHM}}$                         & FWHM Transit Duration                        & days                      & $0.13832^{+0.00038}_{-0.00037}$       \\
b                                                     & Transit Impact Parameter                     & ...                       & $0.29^{+0.10}_{-0.13}$          \\
$\delta_{\mbox{\scriptsize S,2.5 $\mu$m}}$            & Blackbody Eclipse Depth at 2.5 $\mu$m        & ppm                       & $0.65^{+0.19}_{-0.15}$              \\
$\delta_{\mbox{\scriptsize S,5.0 $\mu$m}}$            & Blackbody Eclipse Depth at 5.0 $\mu$m        & ppm                       & $23.8^{+3.2}_{-2.7}$              \\
$\delta_{\mbox{\scriptsize S,7.5 $\mu$m}}$            & Blackbody Eclipse Depth at 7.5 $\mu$m        & ppm                       & $69.1^{+6.1}_{-5.3}$           \\
$\rho_{\mbox{\scriptsize p}}$                         & Planetary Density\tbnm{a}                    & g/cm$^{3}$                & $1.50^{+0.55}_{-0.35}$           \\
$\log{g_{\mbox{\scriptsize p}}}$                      & Surface Gravity\tbnm{a}                      & ...                       & $3.02^{+0.14}_{-0.11}$          \\
$\Theta$                                              & Safronov Number                              & ...                       & $0.0377^{+0.014}_{-0.0086}$       \\
$\langle F \rangle$                                   & Incident Flux                                & 10$^{9}$ erg/(s cm$^{2}$) & $0.0308^{+0.0041}_{-0.0037}$      \\
T$_{\mbox{\scriptsize P}}$                            & Time of Periastron                           & BJD\_TDB                  & $2458736.617288\pm0.000097$   \\
T$_{\mbox{\scriptsize S}}$                            & Time of Eclipse                              & BJD\_TDB                  & $2458740.848893^{+0.00010}_{-0.000099}$    \\
T$_{\mbox{\scriptsize A}}$                            & Time of Ascending Node                       & BJD\_TDB                  & $2458742.96470\pm0.00010$       \\
T$_{\mbox{\scriptsize D}}$                            & Time of Descending Node                      & BJD\_TDB                  & $2458738.733091\pm0.000098$       \\
\mplan$\sin{i}$                                       & Minimum Mass\tbnm{a}                         & \mjup                     & $0.053^{+0.020}_{-0.012}$             \\
\mplan/\mstar                                         & Mass Ratio\tbnm{a}                           & ...                       & $0.000097^{+0.000037}_{-0.000023}$     \\
d/\rstar                                              & Separation at Mid Transit                    & ...                       & $18.66^{+0.56}_{-0.69}$              \\
P$_{\mbox{\scriptsize T}}$                            & A Priori Non-grazing Transit Probability     & ...                       & $0.0510^{+0.0019}_{-0.0015}$          \\
P$_{\mbox{\scriptsize T,G}}$                          & A Priori Transit Probability                 & ...                       & $0.0562^{+0.0022}_{-0.0016}$        \\
u$_{1}$                                               & Linear Limb Darkening Coefficient at 4.5 $\mu$m            & ppm         & $0.052^{+0.061}_{-0.037}$          \\
u$_{2}$                                               & Quadratic Limb Darkening Coefficient at 4.5 $\mu$m         & ppm         & $0.157^{+0.077}_{-0.096}$            \\
A$_{\mbox{\scriptsize T}}$                            & Thermal Emission from Planet                 & ppm                       & $171\pm29$              \\
$\delta_{\mbox{\scriptsize S}}$                       & Secondary Eclipse Depth                      & ppm                       & $171\pm29$              \\
$\sigma^{2}_{\mbox{\scriptsize T}}$                   & Added Variance (Transit)                     & ...                       & $0.0000000986^{+0.0000000038}_{-0.0000000037}$ \\
$\sigma^{2}_{\mbox{\scriptsize S}}$                   & Added Variance (Eclipse)                     & ...                       & $0.0000000210^{+0.0000000034}_{-0.0000000032}$ \\
$F_{0,\mbox{\scriptsize T}}$                          & Baseline Flux (Transit)                      & ...                       & $1.000379\pm0.000031$    \\
$M_{0,\mbox{\scriptsize T}}$                          & Linear detrending coeff (Transit)            & ...                       & $0.000218\pm0.000028$     \\
$M_{1,\mbox{\scriptsize T}}$                          & Quadratic detrending coeff (Transit)         & ...                       & $-0.000149\pm0.000036$     \\
$M_{2,\mbox{\scriptsize T}}$                          & Gaussian detrending coeff (Transit)          & ...                       & $0.001638\pm0.000045$   \\
$F_{0,\mbox{\scriptsize S}}$                          & Baseline Flux (Eclipse)                      & ...                       & $0.999891\pm0.000020$     \\
$M_{0,\mbox{\scriptsize S}}$                          & x detrending coeff (Eclipse, smoothed)       & ...                       & $-0.00133\pm0.00014$   \\
$M_{1,\mbox{\scriptsize S}}$                          & x detrending coeff (Eclipse, high-pass)      & ...                       & $-0.00134\pm0.00015$    \\
$M_{2,\mbox{\scriptsize S}}$                          & y detrending coeff (Eclipse, smoothed)       & ...                       & $0.00269\pm0.00033$    \\
$M_{3,\mbox{\scriptsize S}}$                          & y detrending coeff (Eclipse, high-pass)      & ...                       & $0.00069\pm0.00012$     \\
$M_{4,\mbox{\scriptsize S}}$                          & Noise/Pixel detrending coeff (Eclipse, smoothed)            & ...        & $-0.00453^{+0.00099}_{-0.00098}$   \\
$M_{5,\mbox{\scriptsize S}}$                          & Noise/Pixel detrending coeff (Eclipse, high-pass)           & ...        & $-0.00629\pm0.00047$     \\
$M_{6,\mbox{\scriptsize S}}$                          & $FWHM_{x}$ detrending coeff (Eclipse, smoothed)             & ...        & $-0.00034\pm0.00027$     \\
$M_{7,\mbox{\scriptsize S}}$                          & $FWHM_{x}$ detrending coeff (Eclipse, high-pass)            & ...        & $0.00042\pm0.00036$    \\
$M_{8,\mbox{\scriptsize S}}$                          & $FWHM_{y}$ detrending coeff (Eclipse, smoothed)             & ...        & $0.00250^{+0.00051}_{-0.00052}$   \\
$M_{9,\mbox{\scriptsize S}}$                          & $FWHM_{y}$ detrending coeff (Eclipse, high-pass)            & ...        & $0.00457\pm0.00033$   \\
$M_{10,\mbox{\scriptsize S}}$                         & Sky detrending coeff (Eclipse, smoothed)     & ...                       & $0.00032^{+0.00012}_{-0.00013}$    \\
$M_{11,\mbox{\scriptsize S}}$                         & Sky detrending coeff (Eclipse, high-pass)    & ...                       & $-0.000141^{+0.000025}_{-0.000026}$    \\
$M_{12,\mbox{\scriptsize S}}$                         & Linear detrending coeff (Eclipse)            & ...                       & $0.00084\pm0.00011$     \\
$M_{13,\mbox{\scriptsize S}}$                         & Quadratic detrending coeff (Eclipse)         & ...                       & $0.000454^{+0.000044}_{-0.000045}$   \\
\enddata
\tablecomments{See Table 3 in \citet{eastman2019} for a detailed description of all parameters.}
\tablenotetext{a}{Uses measured radius and estimated mass from \citet{chenkipping2017}}
\tablenotetext{b}{Time of Conjunction is commonly reported as the "transit time".}
\tablenotetext{c}{Time of Minimum Projected Separation is a more correct "transit time".}
\tablenotetext{d}{Optimal Time of Conjunction minimizes the covariance between \tc and Period.}
\tablenotetext{e}{Assumes no albedo and perfect redistribution.}
\end{deluxetable*}

\section{Discussion}\label{sec:discussion}

In this section we discuss the implications of our results and explore possible explanations for the observed planetary thermal emission. In ${\S}$\ref{sec:posteriors} we consider the impact of eclipse depth priors on the recovered eclipse depth. In ${\S}$\ref{sec:temp} we investigate if the observed dayside temperature could be explained by inefficient heat transport or gravitational contraction. In ${\S}$\ref{sec:evolution}, we present formation and evolution simulations. In ${\S}$\ref{sec:pulsations} and ${\S}$\ref{sec:flares}, we address the possibility of stellar pulsations or other activity as an explanation for our results. In ${\S}$\ref{sec:alt}, we discuss some other possible alternative explanations for the observed thermal eclipse depth. Finally, in ${\S}$\ref{sect:detrendingpalooza} we discuss our choice to employ smoothed and high-pass filtered systematic trend time-series.

\subsection{Assessment of the Eclipse Depth Prior on Eclipse Posteriors}\label{sec:posteriors}

\begin{figure*}
    \centering
    \includegraphics[width=180mm]{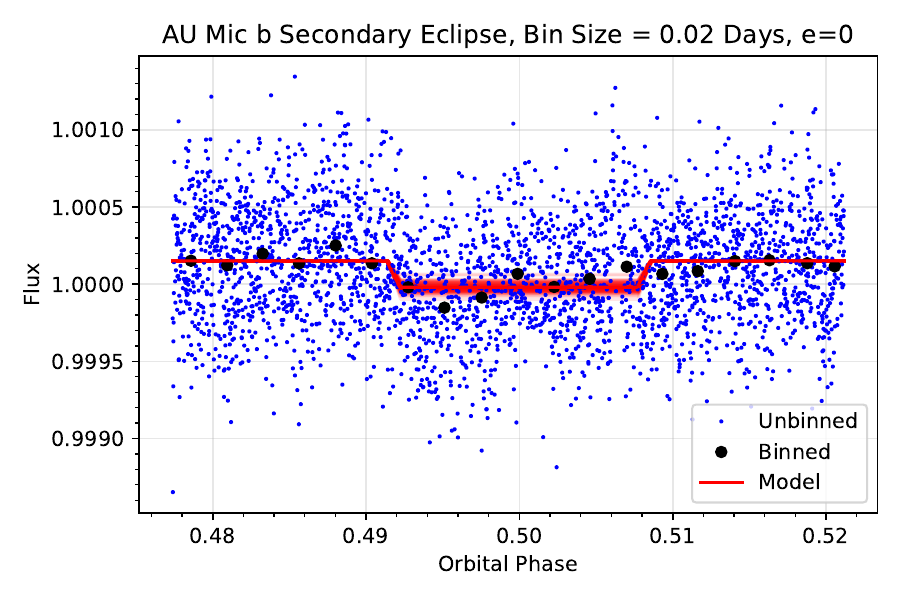}
    \caption{The final \exofast detrended and modeled secondary eclipse light curve. Detrended and phased \spitzer\ data are in blue, data binned to 0.02 days or 0.00236 in phase are in black. The solid red line is the best-fit \exofast model. Transparent red lines are models generated from 100 randomly selected \exofast chains to represent the posterior distribution.}
    \label{fig:binnedmodel}
\end{figure*}

\begin{figure*}
    \centering
    \includegraphics[width=180mm]{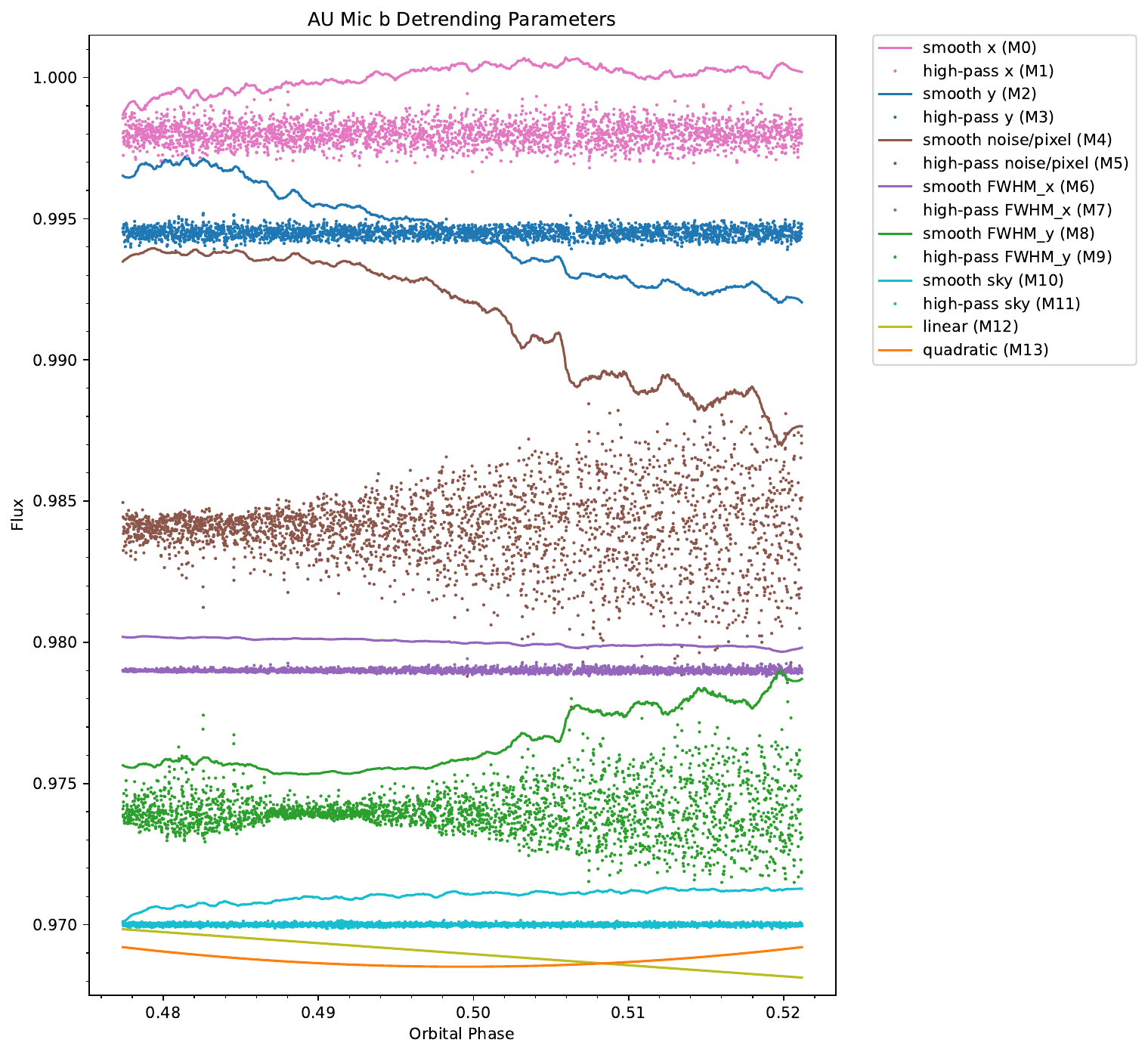}
    \caption{The systematic trend time-series from the \spitzer secondary eclipse light curve of the applied detrending parameters multiplied by their median posterior coefficients and vertically offset. Lines correspond to the smoothed trends, and points represent the high-pass filtered trends.}
    \label{fig:detrend}
\end{figure*}

In this subsection, we assess the impact of the prior for eclipse depth in the recovered eclipse depth posteriors from our MCMC analysis. From our main analysis, the posterior unexpectedly and strongly favors a robust detection of the eclipse at 171$\pm{29}$ ppm, much deeper than expected from geometric and planet formation reasoning. In Fig. \ref{fig:binnedmodel}, we present the final detrended and modeled light curve, and in Fig. \ref{fig:detrend} we show the trend time-series multiplied by their median posterior coefficients to visualize for scale how the detrending time-series are contributing to the final light curve. The strongest contribution after detrending comes from the noise per pixel time-series, both smoothed and high-pass filtered, followed by high-pass filtered \texttt{$FWHM_y$} and smoothed \texttt{$y$}-position. None of the systematic trend time-series multiply together to produce the final eclipse shape present in the data, nor do they exhibit changes near where the ingress and egress are modeled in our data.

We next repeat the MCMC analysis with tighter priors on the eclipse depth that more strongly favor a non-detection — Gaussian priors centered on 0 ppm with widths of 50, 25, and 10 ppm, respectively. For these other Gaussian priors, we obtain depth posteriors of 129$\pm$25, 74$\pm$19, and 18.8$\pm$9.6, respectively. In all cases, the posteriors favor detections and reject the null hypothesis of a non-detection at roughly 5, 4, and 2-$\sigma$, respectively. There is a dependence on the recovered posterior median eclipse depth on the narrowness of the prior, which is to be expected: with priors increasingly biased towards a non-detection, the recovered posterior depths should get weaker and less statistically significant. To put the statistical significance of this eclipse detection in perspective, a 50 ppm width prior centered on zero corresponds to a $1/e^{(-(171^2/50^2))}\sim10^{5}$ probability ratio bias favoring a non-detection over a 171 ppm detection, and yet our analysis still recovers a 129 ppm posterior median. For the 25 ppm (or 10 ppm) width prior, a 171 ppm detection is disfavored by $\sim10^{20}$ ($\sim10^{126}$), resulting in posteriors favoring shallower eclipses. Regardless, the detection remains unambiguous even with these increasingly biased priors favoring a non-detection. The eclipse depth posteriors are shown in Figure \ref{fig:priorwidths} for these prior scenarios, with the final posteriors for the accepted \exofast run in the Appendix. For all priors considered, the stellar and primary transit parameter posteriors remain consistent.

\begin{figure*}
    \includegraphics[width=0.5\linewidth]{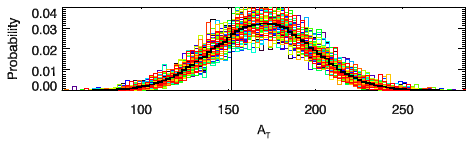}
    \includegraphics[width=0.5\linewidth]{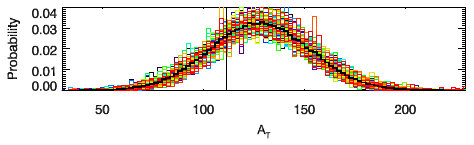}
    \includegraphics[width=0.5\linewidth]{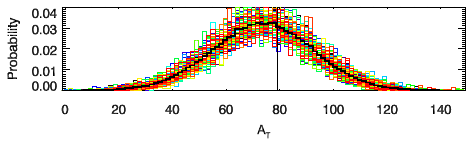}
    \includegraphics[width=0.5\linewidth]{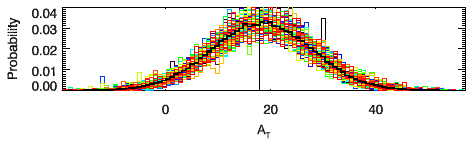}
    \caption{\exofast generated MCMC posterior probability distribution functions for AU Mic b provided different starting prior widths for thermal emission. \textit{Upper Left}: The uninformed case, using a uniform prior with bounds between 0-1000 ppm. \textit{Upper Right}: Gaussian prior centered on 0 ppm with width of 50 ppm. \textit{Lower Left}: Gaussian prior centered on 0 ppm with width of 25 ppm. \textit{Lower Right}: Gaussian prior centered on 0 ppm with width of 10 ppm.}
    \label{fig:priorwidths}
\end{figure*}

\subsection{Dayside Temperature}\label{sec:temp}

\begin{figure*}
    \centering
    \includegraphics{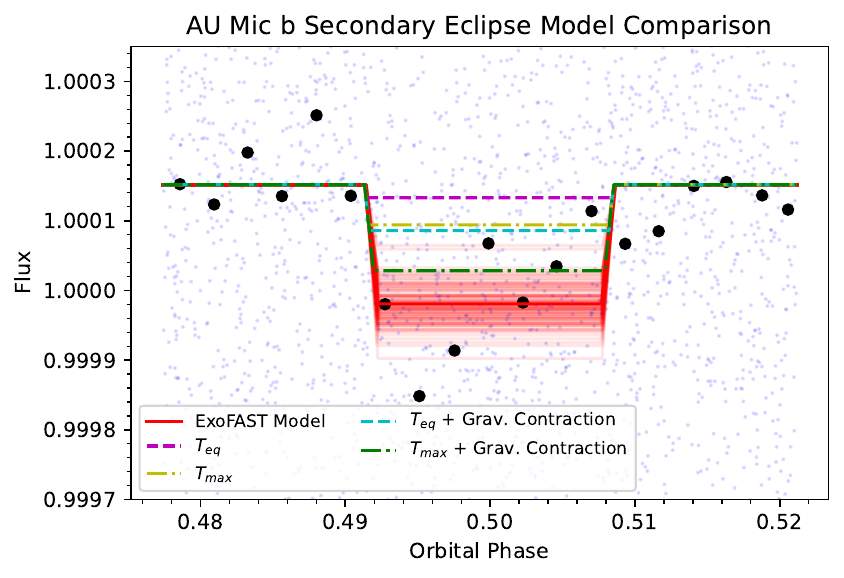}
    \caption{The final phased \spitzer light curve after detrending and modeling, overlaid with models at eclipse depths corresponding to various scenarios. Detrended data are in blue and data binned to 0.02 days are in black. The solid red line is the best-fit \exofast model. Transparent red lines are models generated from 100 randomly selected \exofast chains to represent the posterior distribution. In magenta is a model of an eclipse depth of 18.6 ppm, corresponding to a dayside temperature equal to the derived planet equilibrium temperature, $T_{eq}=606$ K. Yellow corresponds to the theoretical maximum possible dayside brightness temperature, $T_{max}=769$ K, or eclipse depth of 57.7 ppm. Cyan is an eclipse depth of 65.9 ppm, which corresponds to a dayside brightness temperature of $794$ K, a $188$ K increase from the equilibrium temperature estimated from potential gravitational contraction. Finally, green shows a 123.6 ppm eclipse depth, a combination of the heating from gravitational contraction case on top of the theoretical maximum possible dayside brightness temperature.}
    \label{fig:models}
\end{figure*}

Our observed thermal eclipse depth of 171$\pm{29}$ ppm is much deeper than expected, corresponding to a dayside temperature of 1031$\pm{58}$ K, much hotter than the fitted equilibrium temperature of 606$\pm{19}$ K, which would result in a secondary eclipse depth of only 18.6$\pm2.9$ ppm. This surprising discrepancy could be due to any combination of a variety of factors, including inefficient transport of heat to the night side of the planet, excess energy from gravitational contraction, transiting circumstellar dust clouds, or the presence of some circumplanetary material around \aumic b. The last possibility is beyond the scope of this work and will be further explored in a future work. Figure \ref{fig:models} compares our observed eclipse to models at the theoretical maximum dayside brightness temperature and including a realistic level of heat from gravitational contraction, and highlights the discrepancy between these possibilities and what we observe.

Following \citep{Xue2024}, we can compare the determined dayside brightness temperature with the maximum possible dayside brightness temperature, $T_{max}$, by

\begin{equation}
\begin{aligned}
  T_{dayside} &= T_{max} \cdot \mathcal{R} \\
              &= \left(\frac{2}{3}\right)^\frac{1}{4} \cdot \frac{T_{eff}}{\sqrt{a/R_*}} \cdot \mathcal{R} \\
\end{aligned} 
\end{equation}

where the temperature scaling factor $\mathcal{R}$ is defined as
\begin{equation}
\mathcal{R} = \left(\frac{2}{3}\right)^{-\frac{1}{4}} \cdot (1 - A_B)^\frac{1}{4} \cdot (\frac{2}{3} - \frac{5}{12}\epsilon)^\frac{1}{4} \\
\end{equation}
where $A_B$ is Bond albedo and $\epsilon$ is the heat circularization efficiency. When $A_B=0$ and $\epsilon=0$, $\mathcal{R}=1$ and $T_{dayside}$ is the maximum possible dayside brightness temperature, $T_{max}=769\pm49$ K. This shows that the high dayside temperature cannot be explained solely by inefficient heat redistribution.

\subsubsection{Gravitational Contraction - Order of Magnitude Considerations}

Making the simplifying assumption that all of the excess energy between the measured dayside temperature and the expected equilibrium temperature is an effect of gravitational contraction, we can calculate an upper limit of the gravitational contraction rate of \aumic b using the equation:
\begin{equation}
\frac{dU_r}{dt} = \frac{-3GM^2}{10R^2} \frac{dR}{dt} \\
\end{equation}
This results in an upper limit of the contraction rate $\frac{dR}{dt}$ of $\sim$0.27 m $s^{-1}$. This is far too rapid to be physical because contraction at this rate would imply the planet would shrink to a zero radius in only $\sim$3 years. This approach also ignores any potential contributions from chemical energy, nuclear radiative energy, or planetary magnetic field driven heating, in addition to any other possible explanations for the increased dayside temperature.

Conversely, working backwards from more realistic contraction rates, we can estimate the resulting excess energy from gravitational contraction and compare it to the observed dayside temperature. Assuming \aumic b formed from a cloud of material with a radius of 1 AU over the estimated age of the system of 22 Myr results in a contraction rate of 0.2156 mm $s^{-1}$ (three orders of magnitude slower), which results in an excess energy corresponding to an increase in temperature of 188 K, well below the $>400$ K excess observed, illustrating that gravitational potential energy alone is not enough to account for the observed dayside temperature. Although reducing the contraction time does increase the resulting excess temperature, to account for all of the observed excess would require unrealistic contraction timescales of $<1$ Myr.

\begin{figure*}
    \centering
    \begin{tabular}{cc}
    \includegraphics[width=80mm]{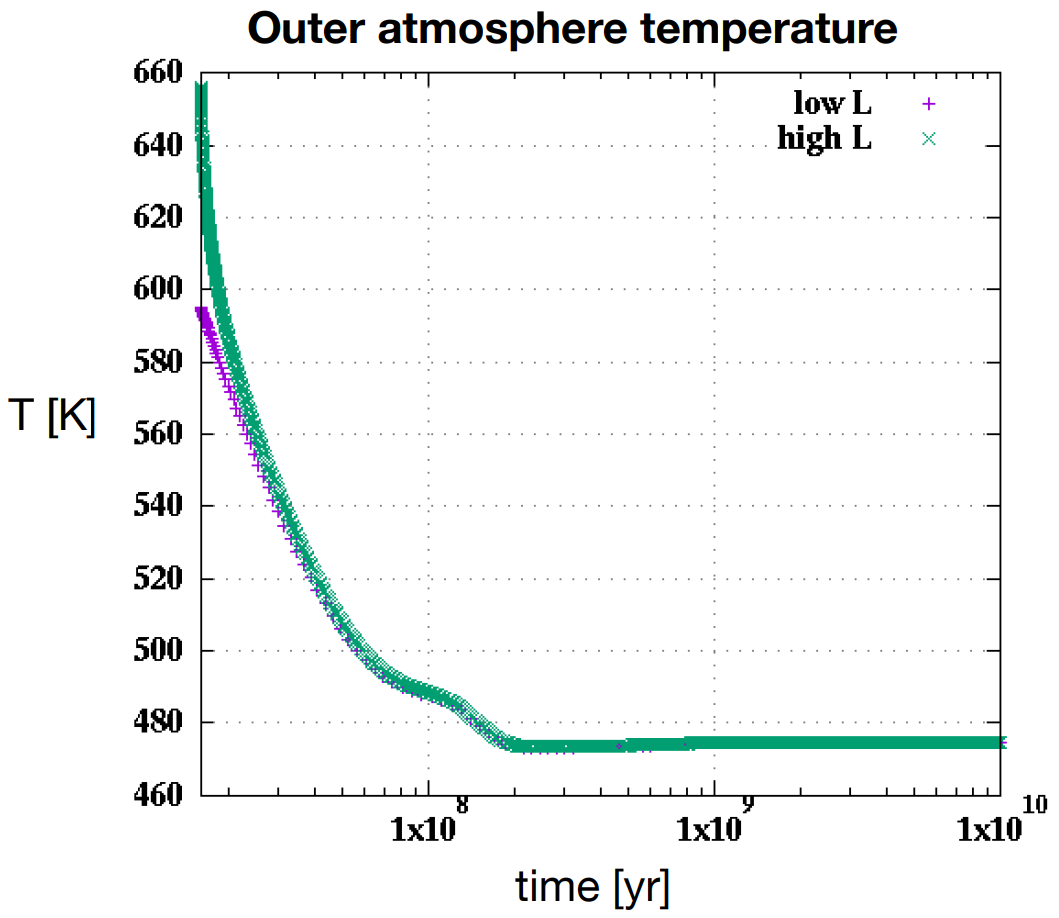} & \includegraphics[width=80mm]{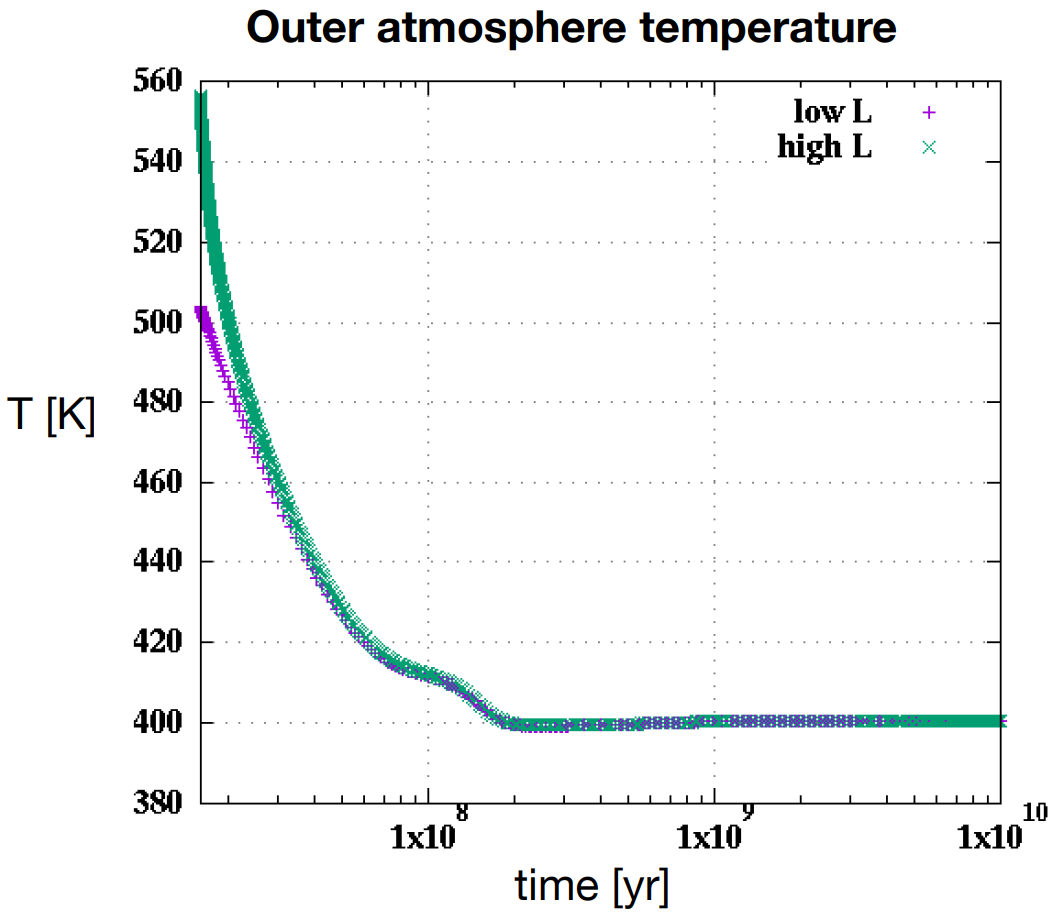} \\
    \end{tabular}
    \caption{Temporal evolution simulations of AU Mic b's atmospheric temperature (which we measured to be 1031$\pm{58}$ K, over 6-$\sigma$ higher than in either simulation scenario), with the high luminosity case in green Xs and the low luminosity case in purple crosses. \textit{Left}: The case of a silicate-iron core with a pure H/He envelope corresponding to a formation scenario inside of the ice line. \textit{Right}: The case of a silicate-iron core with a water-rich envelope of 50\% in mass, representing the maximum predicted for formation beyond the ice line.}
    \label{fig:evolution}
\end{figure*}

\subsection{Detailed Formation Evolution Simulations}\label{sec:evolution}

We generate cooling curves from detailed formation evolution simulations following the framework of \citet{Mordasini2012}. The temporal evolution of the system was simulated considering two potential bulk composition cases: a silicate-iron core with a pure H/He envelope corresponding to a formation scenario inside of the ice line, and a silicate-iron core with a water-rich envelope of 50\% in mass, which represents the maximum predicted for formation beyond the ice line. In all simulations, we assume an Earth-like iron mass fraction in the core of 33\%. We must also guess an internal luminosity of the planet, so here we again consider two extreme cases: a high luminosity case of 100 \ljup, and a low luminosity case of 1 \ljup. In order to mimic the luminosity and effective temperature given in \citep{plavchan2020} while following the evolutionary tracks from \citep{Baraffe2015}, we must set the simulation stellar age to 16 Myr, rather than the estimated consensus age of 22 Myr. After determining the bulk compositions at this age needed to reproduce the observed mass and radius, the simulations were integrated forward to 10 Gyr.

In determining the mass of the pure H/He envelope in order to fit the radius and mass of \aumic b at the given semi-major axis, we find in the high luminosity case a thin envelope of 0.23 \mear atop a 16.87 \mear core, and in the low luminosity case an envelope of 0.8 \mear atop a 16.3 \mear core.

For the case with a H/He and water mixture, we assume that the water and H/He are mixed homogeneously and account for 50\% of the planet's mass, resulting in a silicate-iron core of 8.55 \mear and a volatiles envelope of 8.55 \mear. Instead of adjusting the envelope mass, we adjust the water mass fraction Z to match the observed mass and radius. We find in the high luminosity case an envelope of 90\% water by mass (Z = 0.9) and 10\% H/He. The low luminosity case corresponds to an envelope with Z = 0.78, or 78\% water by mass.

In both bulk composition scenarios, early on in the evolution simulations, the high luminosity cases result in similarly higher temperatures compared to the low luminosity cases, though in all cases these temperatures are much lower than the observed dayside temperature of 1031 K. The temporal evolution of the atmospheric temperatures in both scenarios are shown in Figure \ref{fig:evolution}.

\subsection{Stellar Pulsations}\label{sec:pulsations}

Due to the presence of correlated noise in the \spitzer light curve and the unexpectedly large depth of secondary eclipse, we also investigated the possibility of stellar oscillation pulsations as an explanation for the signal in the secondary eclipse light curve. However, to date there are no confirmed detections of stellar pulsations in M dwarfs, and amplitudes from convective processes like granulation and solar-like oscillations are too small to be responsible for the observed signal. Typical amplitudes for such oscillations are expected to be in the range of 1-10 $\mu$mags and on timescales of only a few minutes \citep{Rodriguez2016, Rodriguez-Lopez2019}. To first order, the frequency at which the oscillations have maximum power, $\nu_{max}$, should scale with stellar mass and radius as follows \citep{Kjeldsen1995}:
\begin{equation}
\nu_{max}=\frac{M/M_{\sun}}{(R/R_{\sun})^2\sqrt{T_{eff}/T_{eff, \sun}}}\nu_{max, \sun}
\end{equation}
For \aumic, this corresponds to a period of 4.9 minutes, which is too short to correspond to the correlated noise seen in the \spitzer light curve.

There are theories for alternative pulsation mechanisms in pre-main sequence stars with amplitudes and timescales that could be a plausible explanation for the observed variability \citep{Steindl2021}. Per Figure 11 of \citet{Steindl2021}, \aumic falls near the predicted instability region for g-modes. However, any such pulsations should have larger amplitudes in the visible than in the infrared. Since we do not see similar variability with greater amplitude on the order of $\sim$0.6 ppt in the \tess light curve, we conclude that the photometric thermal variation we are attributing to the thermal eclipse is likely not of stellar pulsation origin.

\subsection{Flares and other Stellar Activity}\label{sec:flares}

Our analyses rely on the reasonable and supported assumptions of the stellar spin -- planet orbit commensurability ratio and stability of stellar activity between \tess cycles 1 and 3 observations of the AU Mic system, as reviewed in this work. AU Mic is a very active flare star, so it remains feasible that an unlucky combination of flares or abrupt and rapid changes in the rotational modulation of activity break our assumptions in our stellar activity model for the eclipse observation. 

For regular (not abrupt) changes in rotational modulation on the timescale of the eclipse, we include a quadratic model as a systematic trend in our model. We allow for positive and negative sign on the curvature of the quadratic but find the posteriors do not contribute substantially to our light curve model and are not adequate to explain the observed eclipse.

\begin{figure*}
     \centering
    \begin{tabular}{cc}
    \includegraphics[width=80mm]{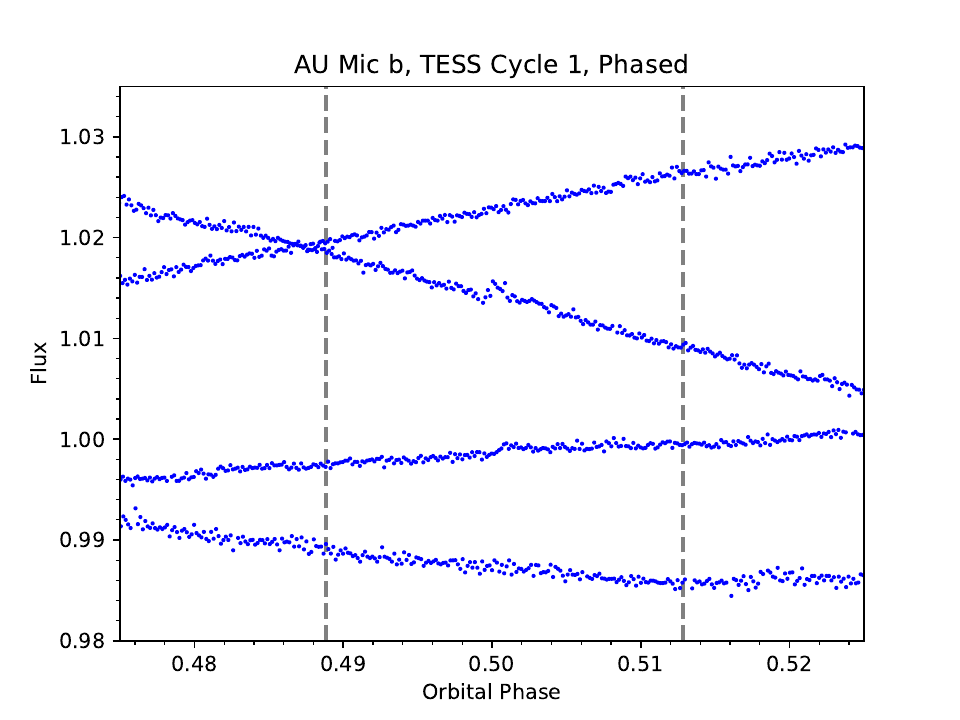} & \includegraphics[width=80mm]{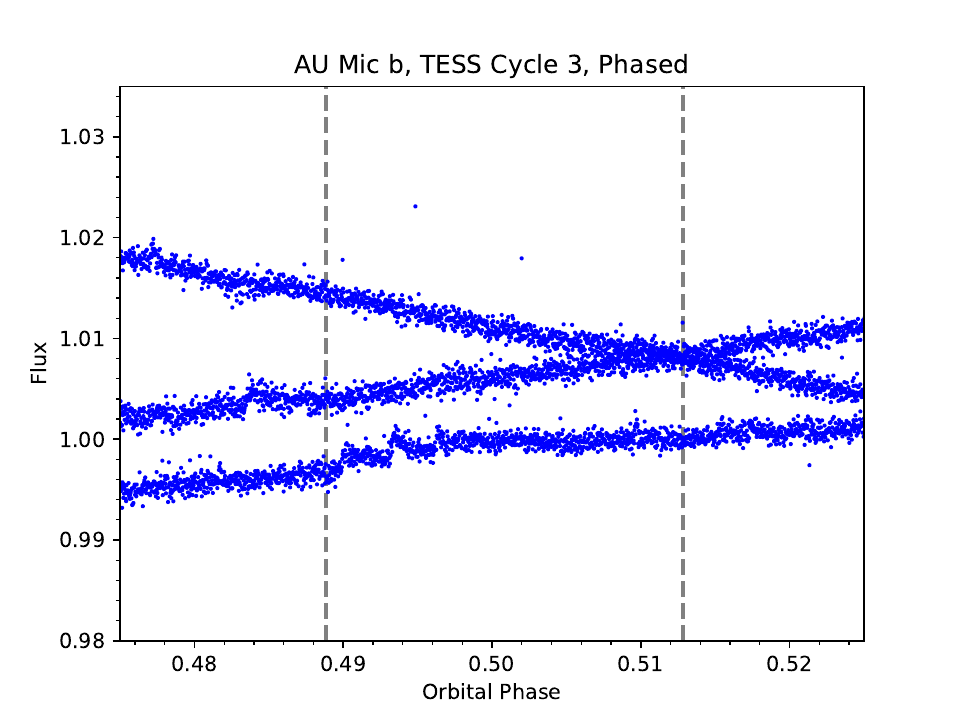} \\
    \end{tabular}
    \caption{Portions of the observed \tess Cycle 1 (\textit{left}) and Cycle 3 (\textit{right}) light curves around modeled times of eclipse as a function of orbital phase. No vertical offsets are applied, and the large-scale variability is from the rotational modulation of stellar activity. The vertical dashed lines correspond to the time of ingress and egress for the observed \spitzer secondary eclipse. The figure illustrates the relative white light flare occurrence rate with respect to the observed eclipse duration.}
    \label{fig:flaresfig}
\end{figure*}

For rapid or abrupt changes such as produced by flares, we explore the flare occurrence rate as a function of flare energy was analyzed in \citet{Gilbert2022} in their Figure 5, who find an asymptote at low flare energy/amplitude occurrence rates of 5.5 flares/day. Given the transit duration of 0.2 days, this corresponds to an average expectation of $\sim$1.2 flares per eclipse. Thus, although unlikely, it remains possible that a complex flare morphology could serendipitously mimic or alter the observed eclipse given the high flare rate for the AU Mic host star. In Figure \ref{fig:flaresfig}, we present the white-light \tess light curves for all eclipse times, and we count 5 flares across 7 eclipse times, consistent with \citet{Gilbert2022} and unlikely to explain our observed eclipse depth and duration. 

\subsection{Discussion Summary: How to Explain the Observed Thermal Eclipse Depth of AU Mic b? Possible Alternatives}\label{sec:alt}

In the preceding subsections, we have discussed a number of astrophysical explanations for the observed thermal eclipse depth and duration for AU Mic b, which imply a very hot dayside temperature. We have reasoned that we can rule out stellar oscillation pulsations, gravitational contraction and evolution and inefficient heat transport between the day and night sides of AU Mic b as viable explanations for our detected apparent dayside temperature of AU Mic b, which is excessively hotter than expected from equilibrium and formation models. The slow rotational modulation of stellar activity is captured in our model and does not explain the observed eclipse. Finally, while a complex flare morphology is not ruled out, a complex flare or rapid brightness change from stellar activity is unlikely to explain the observed eclipse.

One additional explanation that we cannot rule out is that we have additional instrumental systematics in the \spitzer light curve that we have not adequately accounted for in our systematic detrending. AU Mic is bright and relatively red compared to many staring mode observations that were made of transiting exoplanets. However, prior analyses throughout the literature have not reported such a similar systematic present in \spitzer light curves.

In order to definitively exclude stellar activity and instrument systematics as the explanation for the abnormally deep thermal eclipse, future spectroscopic thermal eclipse observations with \textit{JWST} are necessary to confirm our result with \spitzer. If in the future we are able to definitively exclude stellar activity and instrument systematics as the source of our observed thermal eclipse, then we must consider more exotic possibilities to explain the large depth and implied high temperature. First, AU Mic is well known to have an edge-on debris disk geometry existing at semi-major axes $>$35 au \citep{Strubbe2006}, and it is entirely plausible that a planet-sized cloud of circumstellar material could have transited AU Mic at the same time as the predicted thermal eclipse. Again, spectroscopic thermal eclipse observations with \textit{JWST} would help discern this possibility.

Finally, circumplanetary material orbiting AU Mic b could potentially increase the eclipse depth in a manner consistent with what has been observed, resulting in a deeper eclipse than expected from AU Mic b alone. Detailed analyses are required to validate this hypothesis and beyond the scope of this work. No such eclipse observations have been observed to date for another exoplanet, but the transmission spectrum broadband spectral slope of the similarly young and warm Neptune-sized K2-33 b suggests a similar scenario \citep{Ohno2022a, Ohno2022b}. Further, 1SWASP J140747.93-394 542.6 was observed to have what is thought to be a circumplanetary ring system seen in transit over the course of 56-days at a much longer as-of-yet unknown orbital period \citep{Mamajek2012}. Circumplanetary material would also potentially produce detectable signatures in the primary transits of AU Mic b, but no such features have been comprehensively identified or analyzed to date, and such analyses are complicated by the complex stellar activity and flares from the host star. Future \textit{JWST} spectroscopic transit observations again could refute or support this hypothesis.

\subsection{The Importance of Filtering/Smoothing of Systematic Trends}
\label{sect:detrendingpalooza}
In this sub-section, we review the motivation and implications of our choice to smooth the systematic trend time-series with a Savitsky-Golay filter with a time width of 15 minutes as well as incorporating a high-pass filter of trends for the eclipse observation prior to running the MCMC via \exofast. This sub-section presents a temporal accounting at how we arrived at this choice of modeling.

\begin{figure*}
    \centering
    \includegraphics[width=200mm]{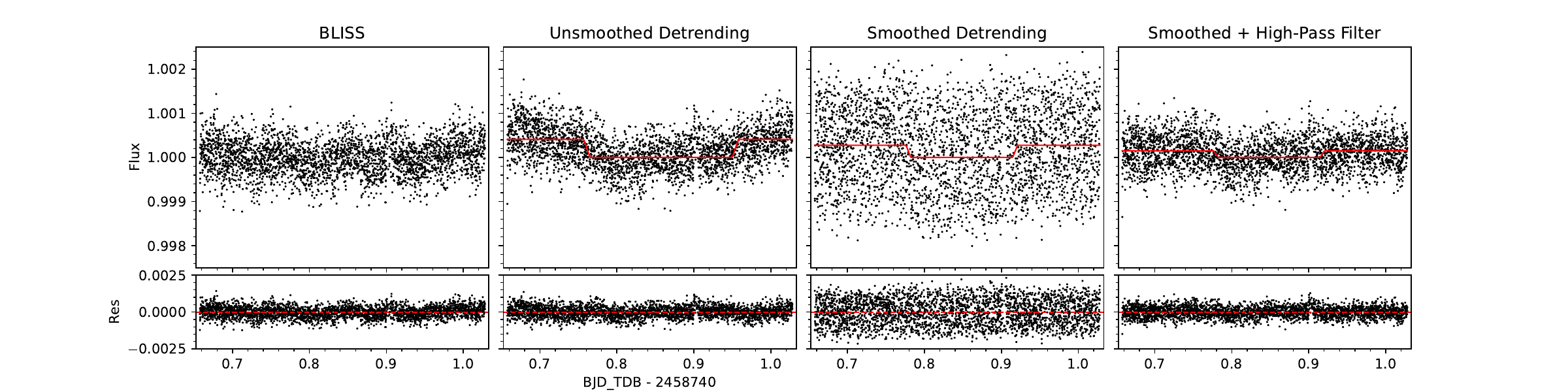}
    \caption{Plots of the \spitzer secondary eclipse light curve and residuals using different detrending methods. The red lines indicate the best-fit eclipse models from \exofast. \textit{Left}: The BLISS-detrended data, assuming zero eclipse. \textit{Center Left}: The raw data after being modeled with \exofast and the detrended by the unsmoothed detrending time-series. \textit{Center Right}: The undetrended \spitzer data after \exofast modeling with smoothed detrending parameters. \textit{Right}: The undetrended \spitzer data after \exofast modeling with both smoothed and high-pass filtered detrending parameters.}
    \label{fig:BLISS-comp}
\end{figure*}

\citet{ingalls2012} finds that BLISS slightly outperforms PLD in both accuracy and reliability in recovered exoplanet parameters, and offers the most robust detrending for Spitzer staring mode observations. \citet{Bell2021} finds that BLISS performs best in most, but not all, of the light curve examples they explore. Our initial analyses of the AU Mic eclipse observation thus started with using BLISS to detrend the eclipse light curve assuming no eclipse was present. We then followed with a secondary detrending with ExoFASTv2 in series, including the eclipse model and all systematic trend time-series to assess their impact on any recovered eclipse depth. This resulted in a detected eclipse with a higher $\sim 10-\sigma$ significance and a similar, but somewhat deeper, eclipse depth to the results presented herein, but as shown in Figure \ref{fig:BLISS-comp} several systematics remained apparent by eye in the light-curve and residuals including a flare-like event. Further, the eclipse duration was anomalously long, implying an eccentricity that was inconsistent with TTV models and requiring some alternative explanations \citep{Wittrock2022,Wittrock2023}. This approach also did not provide a robust assessment of the combined contributions of the systematic time-series to the detrending in the analysis.

\begin{figure*}
    \includegraphics[width=0.5\linewidth]{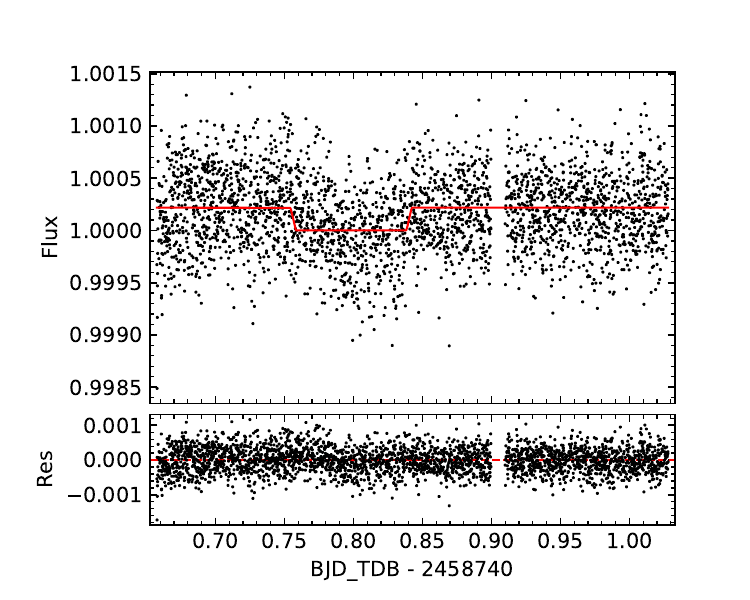}
    \caption{An example of one of the \exofast solutions when modeling with the raw aperture photometry data without smoothing any detrending parameters, an eccentric case with an anomalously short eclipse duration.}
    \label{fig:shorteclipse}
\end{figure*}

In light of the above, we undertook a second approach to the detrending, this time using only \exofast working from the raw aperture-photometry light curve in a single detrending step while jointly modeling the eclipse; e.g. not using BLISS first. In this scenario, the flare-like event remained, but the modeling converged on a series of different solutions, including an eccentric case with an anomalously short eclipse duration inconsistent with the TTVs and dynamical stability considerations, and with visible red-noise systematics remaining in the eclipse residuals as shown in Figure \ref{fig:shorteclipse} \citep{Wittrock2023,Szabo2021}. This approach did however enable us to assess the different contributions of the systematic time-series to the overall detrending.

\begin{figure*}
    \centering
    \includegraphics[width=\textwidth]{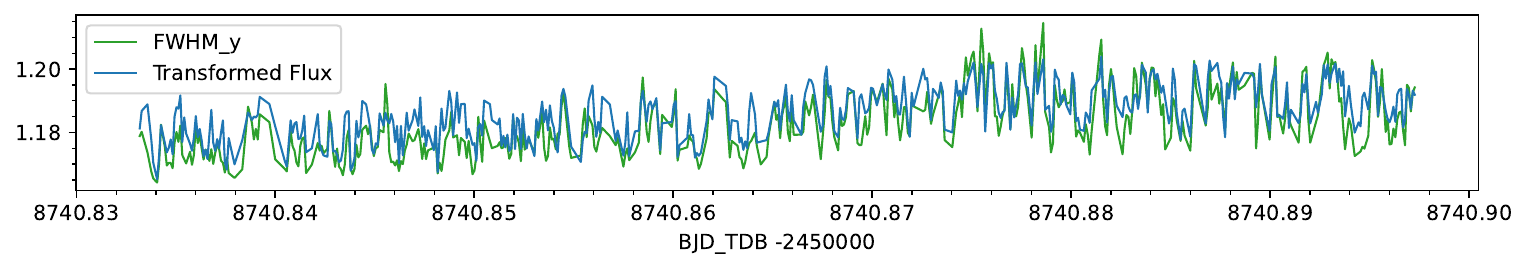}
    \caption{A section of the \spitzer eclipse light curve flux scaled and shifted, illustrating the correlation with the high-frequency noise in the FWHM\_y trend time-series. The flux was arbitrarily scaled and shifted in order to overlap with the trend time-series, using the equation: $scaled\:flux =-7\times(flux-1)+1.18$.}
    \label{fig:high-f-noise}
\end{figure*}

\begin{figure*}
    \centering
    \includegraphics[width=0.3\linewidth]{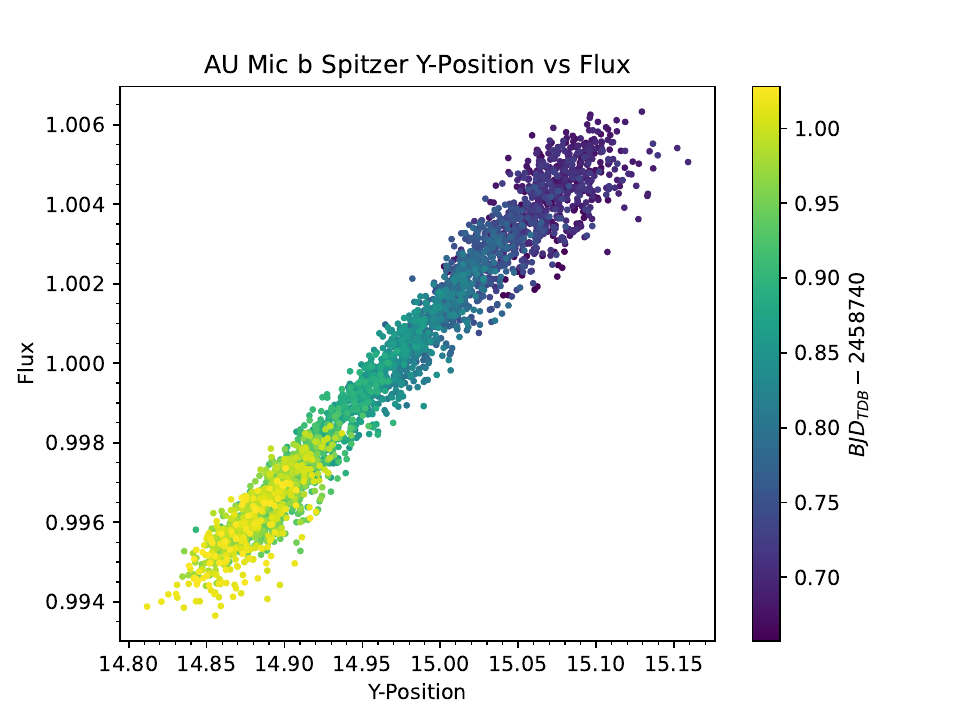}
    \includegraphics[width=0.3\linewidth]{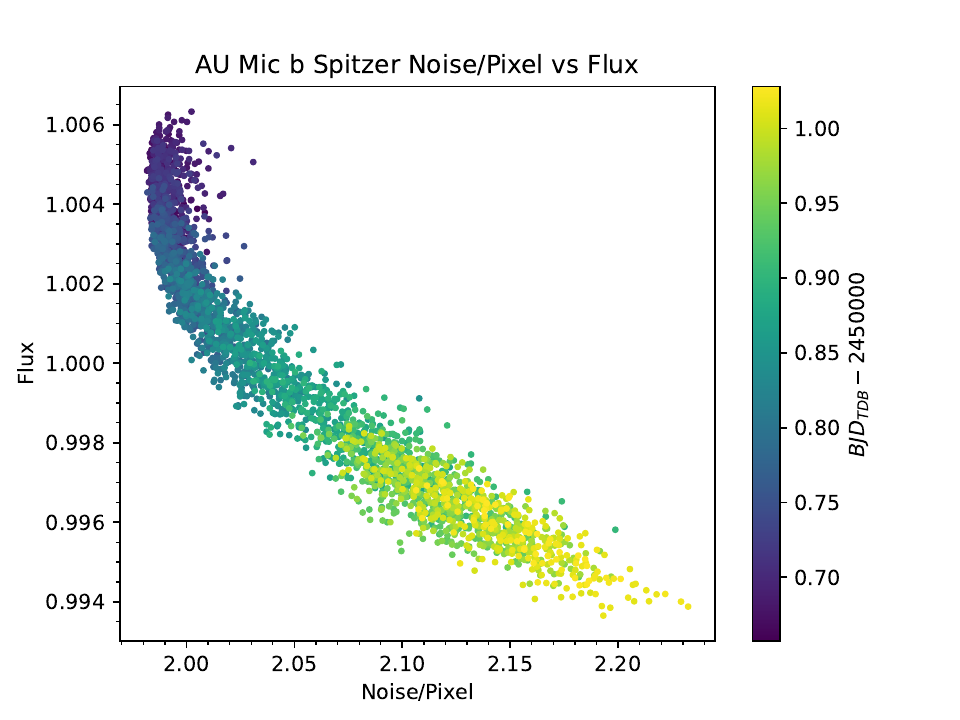}
    \includegraphics[width=0.3\linewidth]{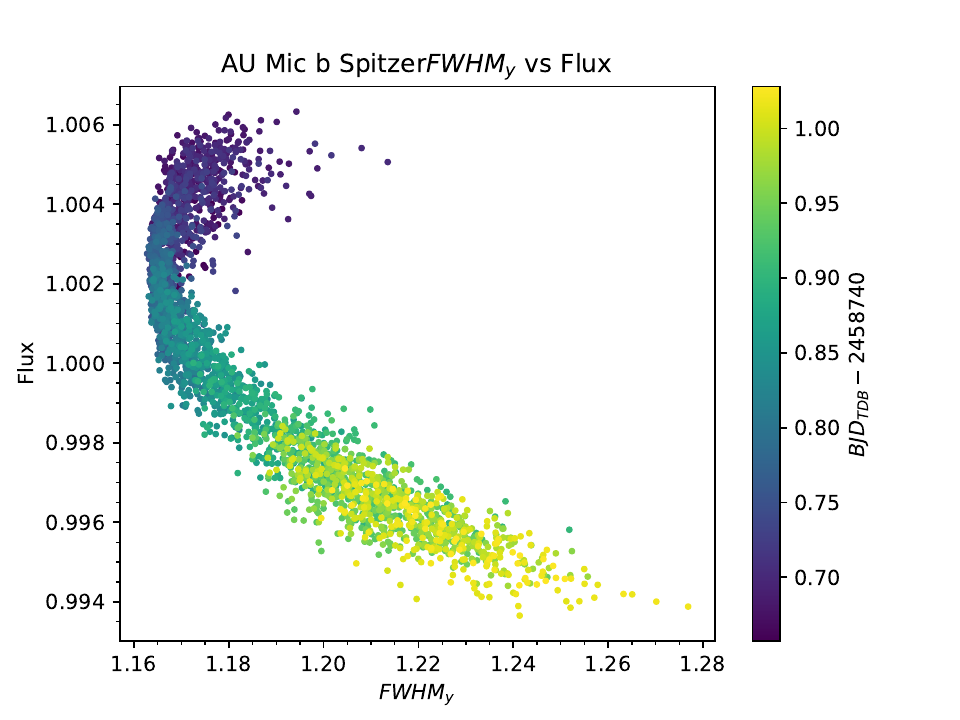}
    \includegraphics[width=0.3\linewidth]{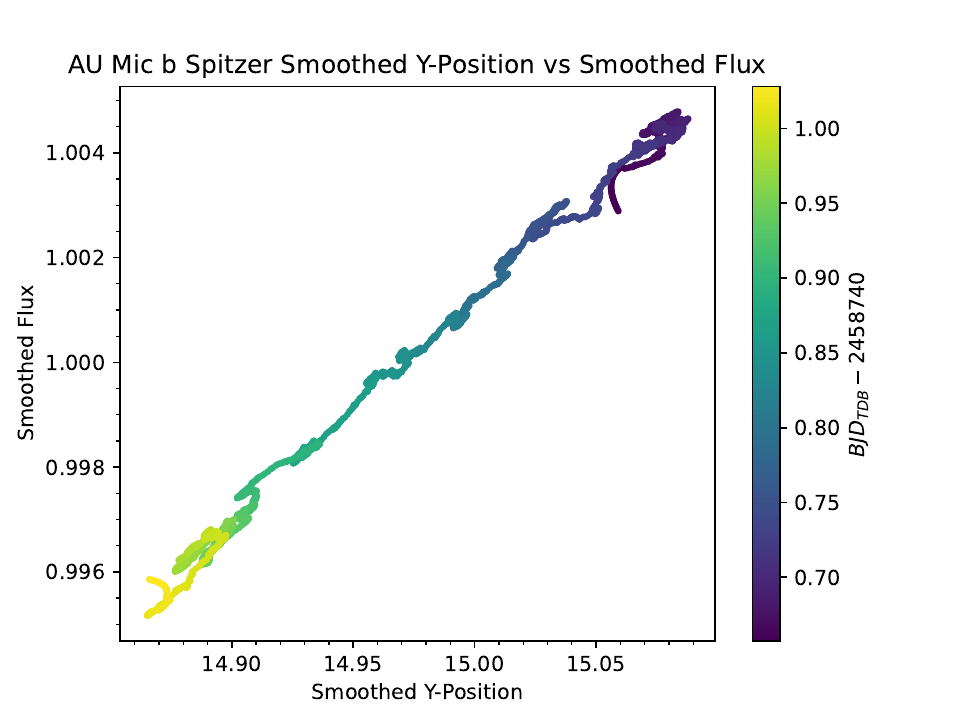}
    \includegraphics[width=0.3\linewidth]{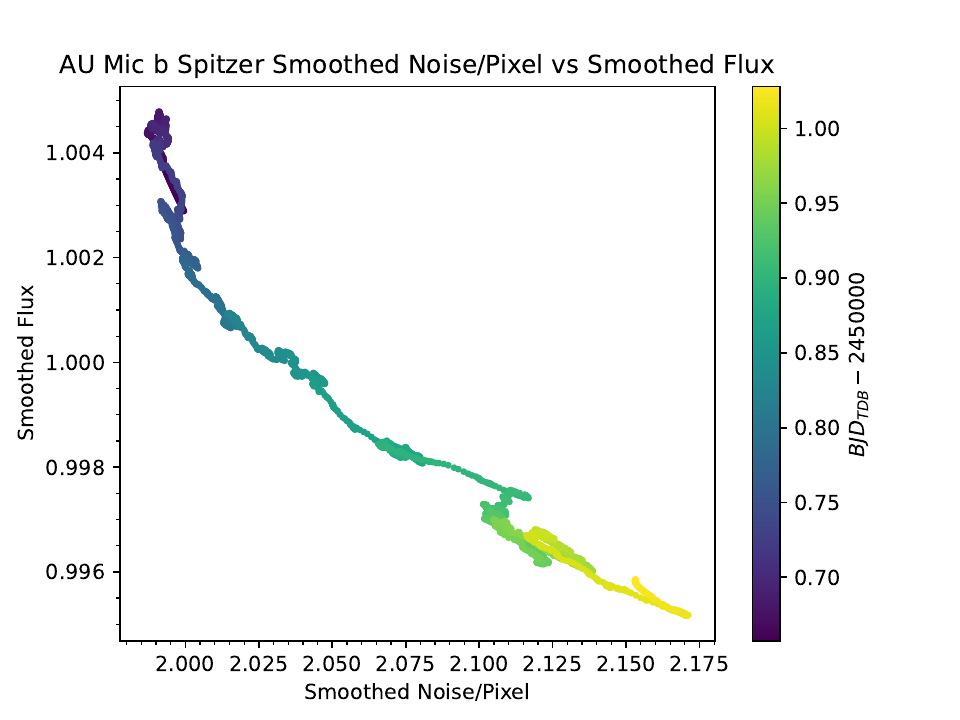}
    \includegraphics[width=0.3\linewidth]{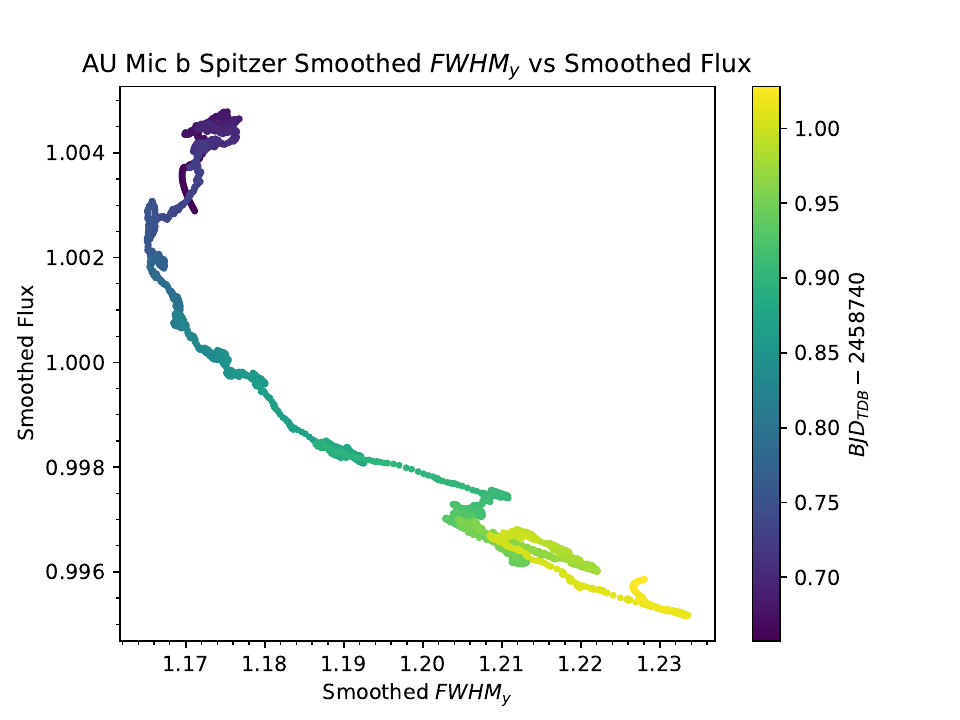}    
    \caption{Plots of the \spitzer secondary eclipse flux versus different detrending parameters, colored by observation time. \textit{Upper Left}: Flux versus Y-position. \textit{Upper Center}: Flux versus noise per pixel. \textit{Upper Right}: Flux versus $FWHM_y$. \textit{Lower Left}: Smoothed flux versus smoothed Y-position. \textit{Lower Center}: Smoothed flux versus smoothed noise per pixel. \textit{Upper Right}: Smoothed flux versus smoothed $FWHM_y$.}
    \label{fig:fluxvdetrends}
\end{figure*}

Befuddled by the inconsistent eclipse detection results from BLISS+\exofast compared to \exofast alone, we performed a manual inspection of the eclipse observation flux and systematic time-series, and discovered a highly-significant linear correlation between flux and y-pixel centroid position that had been previously ignored in both prior detrending methods, with a correlation statistic r-value of 0.96 (Figure \ref{fig:fluxvdetrends}). The prior methods favored detrending primarily with the FWHM$_y$ and noise-per-pixel systematic time-series, which also showed strong correlations with flux, but non-linear in nature (Figure \ref{fig:fluxvdetrends}). This led us to conclude that the optimal detrending model solution should likely be dominated by detrending with the y-pixel centroid position time-series, but that \exofast and BLISS were not converging to this optimal model. We next turned to figuring out why, and visual inspection revealed high-frequency variations in the systematic time-series -- see for example, Figure \ref{fig:high-f-noise} -- that perhaps were preventing convergence of our model on detrending primarily with the y-pixel centroid position that captures all of the prominent low-frequency systematics present in the light curve.

Exoplanet transit light curve systematic detrending tools such as \exofast, PLD, and BLISS used herein, as well as AstroImageJ and others, generally do not account for the noise spectrum of the systematic time series, which can include both shot (instantaneous) noise and correlated noise on longer time-scales \citep{stevenson2012,eastman2019,Collins2017,ingalls2012,Bell2021}. When justified, a correlated noise model can be implemented using a Gaussian Process (GP) to an MCMC \citep[][ and references therein]{Rasmussen2006, Haywood2014, Hogg2018, Liang2020}. However, GPs can be prohibitively computationally expensive for flux time-series \citep[e.g., ][]{plavchan2020}. 

The \spitzer AU Mic eclipse observation systematic time-series show both high-frequency and low-frequency systematic noise, as well as shot noise, all of which are relevant to our analysis. The eclipse ingress and egress are approximately 10 minutes in duration, whereas the known \spitzer systematic timescales include the one-hour long on-board heater cycle \citep{ingalls2012}.

\begin{deluxetable*}{l|c|c|c|c|c|c|c}
    \tablecaption{\label{table:loglikes}\aumic b \exofast eclipse model run comparisons.}
    \tablehead{Run Description & $log(\mathcal{L})$ & $BIC$ & $AIC$ & $\Delta BIC$ & $\Delta AIC$ & $A_T$ (ppm) & $\sigma$}
    \startdata
\parbox{5cm}{\vspace{2pt}\textbf{Final model:} 15-min. smoothed and high-pass trends, circular} & 57040.843 & -113714.13 & -114015.69 & --        & --        & $171\pm29$     & 5.90 \\
Tighter $A_T$ prior\tbnm{*}: $0\pm50$ ppm               & 57036.155 & -113704.75 & -114006.31 & 9.38     & 9.38     & $129\pm25$    & 5.16     \\
Tighter $A_T$ prior\tbnm{*}: $0\pm25$ ppm                                     & 57030.543 & -113693.52 & -113995.09 & 20.61    & 20.60    & $74\pm19$      & 3.89 \\
Tighter $A_T$ prior\tbnm{*}: $0\pm10$ ppm                                     & 57024.424 & -113681.29 & -113982.85 & 32.84    & 32.84    & $18.8\pm9.6$   & 1.96 \\
Unsmoothed trends, circular     & 56994.306 & -113696.02 & -113934.61 & 18.11    & 81.08    & $129\pm22$     & 5.86 \\
BLISS detrended data, circular      & 56958.640 & -113624.74 & -113863.28 & 89.39    & 152.41   & $11.1^{+14}_{-8.1}$ & 0.79 \\
Unsmoothed trends, eccentric & 56657.771 & -112979.93 & -113253.54 & 734.20 & 762.15 & $183\pm23$ & 7.96 \\
5-min. smoothed trends, circular & 54499.047 & -108705.5 & -108944.09 & 5008.63 & 5071.60 & $168\pm55$ & 3.05 \\
15-min. smoothed trends, eccentric    & 54441.350 & -108546.76 & -108820.70 & 5167.37  & 5194.99  & $290^{+210}_{-100}$ & 1.38 \\
15-min. smoothed trends, circular   & 54436.321 & -108580.05 & -108818.64 & 5134.08  & 5197.05  & $236\pm67$     & 3.52 \\
\parbox{5cm}{15-min. smoothed trends, no linear or quadratic trends, circular} & 54429.531 & -108591.26 & -108809.06 & 5122.87 & 5206.63 & $286\pm66$ & 4.33 \\
30-min. smoothed trends, circular & 54414.571&-108536.55&-108775.14&5177.58&5240.55	&$249\pm71$&3.51 \\
Smoothed y-position only, circular  & 54379.044 & -108551.92 & -108718.09 & 5162.21  & 5297.60   & $8.1^{+12}_{-6}$    & 0.68   \\
60-min. smoothed trends, circular    &  54348.263 & -108403.93&	-108642.53&	5310.20&	5373.16&	$316^{+82}_{-84}$&	3.76 \\
\enddata
\tablenotetext{*}{Aside from the $A_T$ prior value, all other parameters are identical to the final model.}
\end{deluxetable*}

The above led us to next adopt low-pass filtering of our systematic time-series as a fundamental change to our modeling approach for the \spitzer eclipse light curve. We explored a variety of smoothing methods and time-scales, arriving at a Savitsky-Golay filter with a width of 15 minutes, achieving consistent results with widths anywhere from 10--30 minutes. This filtered out the highest frequency systematics in the trend time-series, resulting in a flatter baseline light curve without the previously visible red-noise systematics, at the expense of higher instantaneous shot noise and degraded eclipse depth SNR, as shown in Figure \ref{fig:BLISS-comp}. 

We next discovered that the high-frequency flux shot noise is highly correlated with the high-frequency systematic trend variations as illustrated in Figure \ref{fig:high-f-noise}. Therefore it is necessary to retain the high-frequency systematics in our detrending. These high-frequency variations were what had previously determined the trends with the largest contributions to the overall detrending, effectively de-emphasizing the low-frequency systematics in the detrending. Additionally, as seen in Figure \ref{fig:detrend}, the high-frequency variations in the systematic time-series have amplitudes that vary slowly over the baseline of the observations, which is not present in the flux time-series. We thus implemented our final adopted detrending approach as described in ${\S}$\ref{sect:analysis_exofast}. We include both the high-frequency filtered and low-frequency filtered trend time-series as decoupled sets of systematic time-series. While the addition of the high-frequency and low-frequency filtered time-series for a particular systematic is equal to the original systematic time-series (before normalization and modeling the coefficients), this approach enables us to assess the systematic contributions of high and low frequencies independently. This analysis resulted in a larger contribution from the highly correlated low-frequency y-position time-series while still preserving the reduced instantaneous shot noise in the flux from the high-frequency systematic trend variations. 

Table \ref{table:loglikes} compares the results of the various detrending methods employed, illustrating that our final adopted approach of both high-frequency and low-frequency filtering of the systematic trend time-series results in the highest log-likelihood and lowest AIC and BIC among those tested. In fact, the success of this model enables us to strongly rule out the other detrending approaches, with the uninformed prior on $A_T$ still strongly favored over the $A_T=0\pm50$ ppm next best model, the latter of which also produces a detection at similar statistical signicance and slightly reduced amplitude as might be expected.  

AU Mic is uniquely both very bright and red among \spitzer staring mode observation targets, which may have contributed to the necessity and ability (SNR) to decouple high and low-frequencies in systematic time-series and achieve our recovered eclipse detection. We did not attempt, for example, to consider a more sophisticated analysis to normalize the slowly time-varying amplitude of the high-frequency variations in the systematic time-series, which may have yielded further improvements. We also have additional smaller amplitude low-frequency systematics that remain in our resulting light curve, as best illustrated in the second row of Figure \ref{fig:fluxvdetrends} for the low-frequency y-centroid position time-series. However, we do not identify additional systematics, detrending approaches, or astrophysics to model these and defer to future work. Nonetheless, our analysis indicates that there are potential improvements to be achieved by modeling archival \spitzer staring mode observations by adopting more sophisticated detrending methods that take into consideration the frequency domain of systematic time-series. 

Finally, we note that smoothing or altering flux time-series effectively alters the signal-to-noise ratio of flux time-series, and thus artificially alters the relative log-likelihoods of models and model comparisons. However, in our analysis we intentionally did not alter our flux time-series. We only altered our models for the flux time-series by altering our systematic trend time-series, which does not impact the assessment of relative log-likelihoods.

\section{Conclusions} \label{sec:conclusion}

We have conducted \spitzer 4.5 $\mu$m observations of AU Mic b at the predicted time of eclipse. We have analyzed these observations and incorporated a model for the stellar activity at this wavelength from \tess observations. We develop a novel detrending approach for \spitzer staring mode observations that decouple high and low frequency systematic variations to achieve an overall reduction in flux systematics. We have detected a statistically robust event consistent with a thermal eclipse of AU Mic b. The eclipse implies a dayside temperature of T=1031$\pm{58}$ K that is nearly twice as hot as planet formation models, and we confirm a circular orbit and exclude an eccentric model given our data, consistent with TTV analyses. We explore and discuss a range of alternative explanations for our eclipse detection and inferred dayside temperature, including stellar activity, oscillation pulsations, instrumental systematics, inefficient day to night side heat transfer, gravitational contraction estimates, and more intriguing possibilities such as circumplanetary and circumstellar material. We conclude future spectroscopic observations with \textit{JWST} are needed to verify this eclipse detection.

\clearpage
\bibliography{bib_master}{}
\bibliographystyle{aasjournal}

\appendix
\label{sect:appendix}

In this Appendix, we present the \exofast generated MCMC posterior probability distribution functions and corner plot for AU Mic b. \\

\includegraphics[width=\textwidth,page=1,trim={2cm 12.8cm 1cm 5.5cm},clip]{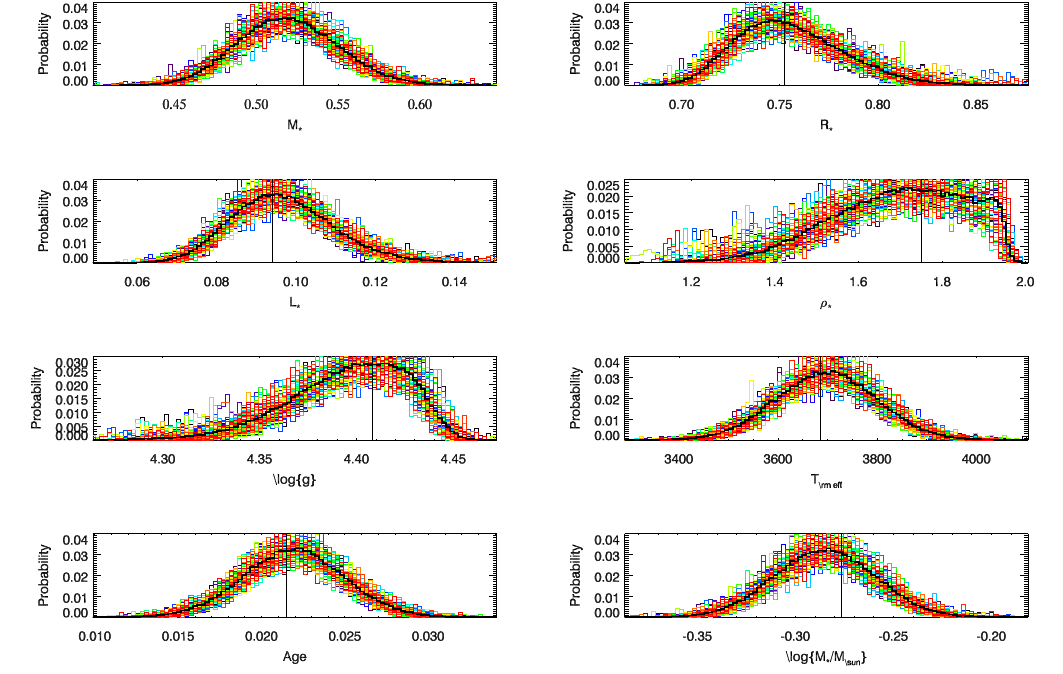}

\includegraphics[width=\textwidth,page=2,trim={2cm 12.8cm 1cm 5.5cm},clip]{AUMic_b.posteriors.pdf}
\includegraphics[width=\textwidth,page=3,trim={2cm 12.8cm 1cm 5.5cm},clip]{AUMic_b.posteriors.pdf} 
\includegraphics[width=\textwidth,page=4,trim={2cm 12.8cm 1cm 5.5cm},clip]{AUMic_b.posteriors.pdf} 
\includegraphics[width=\textwidth,page=5,trim={2cm 12.8cm 1cm 5.5cm},clip]{AUMic_b.posteriors.pdf} 
\includegraphics[width=\textwidth,page=6,trim={2cm 12.8cm 1cm 5.5cm},clip]{AUMic_b.posteriors.pdf} 
\includegraphics[width=\textwidth,page=7,trim={2cm 12.8cm 1cm 5.5cm},clip]{AUMic_b.posteriors.pdf} 
\includegraphics[width=\textwidth,page=8,trim={2cm 12.8cm 1cm 5.5cm},clip]{AUMic_b.posteriors.pdf} 
\includegraphics[width=\textwidth,page=9,trim={2cm 12.8cm 1cm 5.5cm},clip]{AUMic_b.posteriors.pdf} 

\begin{figure*}[ht]
    \centering
    \includegraphics[width=\textwidth,trim={4cm 14cm 7.5cm 4cm},clip]{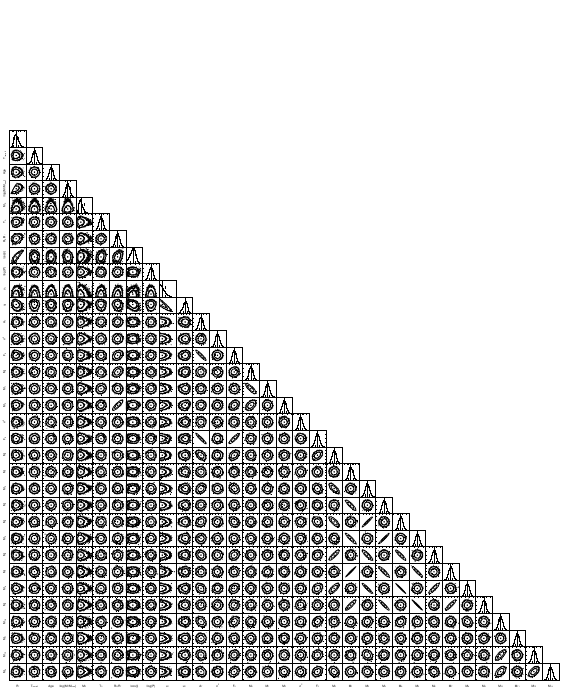}  
    \caption{\exofast generated MCMC corner plot for AU Mic b.}
    \label{fig:corner}
\end{figure*}
\clearpage



\end{document}